\documentclass[12pt]{article}

\usepackage{amssymb}
\usepackage{amsmath}
\usepackage{amscd}
\usepackage{latexsym}
\usepackage{url}

\usepackage{cite}

\usepackage{graphicx}
\usepackage{subfigure}

\newcommand{\be}{\begin{equation}}
\newcommand{\ee}{\end{equation}}

\newcommand{\dlt}{\delta}
\newcommand{\prt}{\partial}

\newcommand{\ep}{\varepsilon}
\newcommand{\al}{\alpha}
\newcommand{\ra}{\rightarrow}
\newcommand{\sgm}{\sigma}

\newcommand{\gm}{\gamma}

\newcommand{\lbd}{\lambda}

\begin{document}

\begin{center}

{\Large{\bf Extreme events in population dynamics with
functional carrying capacity} \\[5mm]

V.I. Yukalov$^{1,2}$, E.P. Yukalova$^{1,3}$, and
D. Sornette$^{1,4}$} \\ [3mm]

{\it
$^1$Department of Management, Technology and Economics, \\
ETH Z\"urich, Swiss Federal Institute of Technology, \\
Z\"urich CH-8032, Switzerland \\ [3mm]

$^2$Bogolubov Laboratory of Theoretical Physics, \\
Joint Institute for Nuclear Research, Dubna 141980, Russia \\ [3mm]

$^3$Laboratory of Information Technologies, \\
Joint Institute for Nuclear Research, Dubna 141980, Russia \\ [3mm]

$^4$Swiss Finance Institute, c/o University of Geneva, \\
40 blvd. Du Pont d'Arve, CH 1211 Geneva 4, Switzerland}

\end{center}

\vskip 2cm

\begin{abstract}

A class of models is introduced describing the evolution 
of population species whose carrying capacities are
functionals of these populations. The functional dependence of
the carrying capacities reflects the fact that the correlations
between populations can be realized not merely through direct
interactions, as in the usual predator-prey Lotka-Volterra
model, but also through the influence of species on the carrying
capacities of each other. This includes the self-influence of
each kind of species on its own carrying capacity with delays. 
Several examples of such evolution equations with functional carrying
capacities are analyzed. The emphasis is given on the conditions
under which the solutions to the equations display extreme events, 
such as finite-time death and finite-time singularity. Any destructive 
action of populations, whether on their own carrying capacity or on 
the carrying capacities of co-existing species, can lead to the
instability of the whole population that is revealed in the form of 
the appearance of extreme events, finite-time extinctions or booms 
followed by crashes.

\end{abstract}

\vskip 1cm



\newpage

\section{Brief survey of population models}

Evolution equations, describing population dynamics, are widely
employed in various branches of biology, ecology, and sociology.
The main forms of such equations are given by the variants of the
predator-prey Lotka-Volterra models. In this paper, we introduce
a novel class of models whose principal feature, making them different
from other models, is the functional dependence of the population
carrying capacities on the population species. This general class
of models allows for different particular realizations
characterizing specific correlations between coexisting species.
The functional dependence of the carrying capacities describes the
mutual influence of species on the carrying capacities of each
other, including the self-influence of each kind of species on
its own capacity. Such a dependence is, both mathematically and
biologically, principally different from the direct interactions
typical of the predator-prey models. Before formulating the general
approach, we give in this section a brief survey of the main known
models of population dynamics. This will allow us to stress the
basic difference of our approach from other models used for
describing the population dynamics in biology, ecology, and
sociology.

\vskip 2mm

(i) {\bf Predator-prey Lotka-Volterra model}

\vskip 2mm

The first model, describing interacting species, one of which is
a predator with population $N_1$, and another is a prey with population $N_2$,
has been the Lotka-Volterra \cite{1,2} model
\be
\label{1}
\frac{dN_1}{dt} = -\gm_1 N_1 + A_{12} N_2 N_1 \; , \qquad
 \frac{dN_2}{dt} = \gm_2 N_2 - A_{21} N_1 N_2 \;  ,
\ee
where all coefficients are positive numbers. It is easy to show
that the solutions to these equations are bound oscillating
functions of time.

\vskip 2mm

(ii) {\bf Predator-prey Kolmogorov model}

\vskip 2mm

The Lotka-Volterra model is a particular case of the predator-prey
Kolmogorov model \cite{3,4} that has the general form
\be
\label{2}
\frac{dN_1}{dt} = f_1(N_1,N_2) N_1 \; , \qquad
\frac{dN_2}{dt} = f_2(N_1,N_2) N_2 \;  ,
\ee
under the conditions
$$
 \frac{\prt f_1}{\prt N_2} \; > \; 0 \; , \qquad
 \frac{\prt f_2}{\prt N_1} \; < \; 0 \;  .
$$
This model is too general and requires specifications
for describing concrete cases.

\vskip 2mm

(iii) {\bf Generalized predator-prey Lotka-Volterra model}

\vskip 2mm

Generalizing the Lotka-Volterra model (\ref{1}) to multiple
species yields the equations
\be
\label{3}
 \frac{dN_i}{dt} = \left ( \gm_i +
\sum_j A_{ij} N_j \right ) N_i \;  ,
\ee
where all coefficients are real numbers \cite{5}. The signs of
the parameters can be different. When all $\gamma_i$'s are
positive, while all $A_{ij}$'s are negative, one gets the
competitive Lotka-Volterra equations whose behavior has been
analyzed in detail in Refs. \cite{6,7,8,9}.

\vskip 2mm

(iv) {\bf Replicator equations}

\vskip 2mm

These equations have the form
\be
\label{4}
\frac{dN_i}{dt} = \left ( f_i - \overline f \right ) N_i \; ,
\ee
where $f_i$ is the species fitnesses and
$$
\overline f \equiv \sum_i f_i N_i
$$
is the average fitness characterizing the whole society \cite{5}.
The species populations are usually assumed to be defined on
a simplex, being normalized to a constant representing the
total fixed population
$$
\sum_i N_i = N = const \;  .
$$
The $n$- dimensional replicator model is equivalent to the
$n-1$-dimensional Lotka-Volterra model (\ref{3}), to which
it can be reduced by a change of variables \cite{5,10}.

\vskip 2mm

(v) {\bf Jacob-Monod equations}

\vskip 2mm

The equations describe not the coexisting species but rather
a single type of species of population $N_1$, like bacteria,
which are fed on the nutrient of amount $N_2$. The nutrient
plays the role of the prey that is getting depleted being
consumed by the feeders and, at the same time, being supplied
into the system from outside according to the supply function
$f(N_2) = \alpha N_2 / (\beta + N_2)$. The equations read as
\be
\label{5}
\frac{dN_1}{dt} = f(N_2) N_1 \; , \qquad
\frac{dN_2}{dt} = -\gm f(N_2) N_1   ,
\ee
with all parameters being positive \cite{11}. As time increases,
$t \ra \infty$, the nutrient becomes depleted, $N_2 \ra 0$,
and the bacteria population reaches the stationary value
$N_1 = N_1(0) + N_2(0)/ \gamma$. 

The Holling equation of second kind \cite{12} takes into account that
predators, in order to consume prey, need to search for it, chase, 
kill, eat, and digest. This is why predators attack not all preys 
but a limited number of them, which saturates to a constant when 
the prey density increases. Mathematically, the Holling equation
is analogous to the Jacob-Monod model. 

\vskip 2mm

(vi) {\bf Verhulst logistic equation}

\vskip 2mm

The well known logistic equation
\be
\label{6}
\frac{dN}{dt} = \gm N \left ( 1 \; - \; \frac{N}{K} \right ) \;  ,
\ee
where all parameters are positive, was suggested by Verhulst \cite{13}.
The constant $K$ is the carrying capacity. The solution to this
equation is the sigmoid function
$$
N(t) =
\frac{N_0 K e^{\gm t} }{K+N_0\left (e^{\gm t} - 1\right )} \; ,
$$
in which $N_0 \equiv N(0)$.

\vskip 2mm

(vii) {\bf Hutchinson delayed logistic equation}

\vskip 2mm

If one interprets the term inside the brackets in Eq. (\ref{6}) as 
an effective reproductive rate, then, as Hutchinson argued \cite{14},
it could be delayed in time, which results in the equation
\be
\label{7}
\frac{dN(t)}{dt} = \gm N(t) \left [ 1 \; - \;
\frac{N(t-\tau)}{K} \right ] \;  ,
\ee
in which $K$ is a fixed carrying capacity. The solution to this
equation gives additional oscillations superimposed on the
logistic curve.

\vskip 2mm

(viii) {\bf Generalized delayed logistic equations}

\vskip 2mm

There are many variants generalizing the delayed logistic
equation (\ref{7}), which can be found in Refs. \cite{15,16,17}. For
example, the multiple-delayed equation
\be
\label{8}
 \frac{dN(t)}{dt} = \gm N(t) \left [ 1 \; - \sum_j \;
\frac{N(t-\tau_j)}{K_j} \right ] \;  ,
\ee
where the carrying capacities $K_j$ are positive constants.
All such equations are usually applied to single-species
systems of population $N$. The multiple carrying capacities
$K_j$ in Eq. (\ref{8}) correspond to different processes of the
same single species. The logistic equations of the above type,
whether with delays or without them, do not describe the
possible coexistence of several species. When such equations
are generalized to the case of several species, one comes back
to the generalized Lotka-Volterra predator-prey model (\ref{3}).

\vskip 2mm

(ix) {\bf Peschel-Mende hyperlogistic equation}

\vskip 2mm

In order to take into account the accelerated growth of population,
occurring, for instance, for the human world population, Peschel and
Mende \cite{18} extended the standard logistic equation by introducing
two additional positive powers $m$ and $n$, getting the equation
\be
\label{8a}
 \frac{dN}{dt} = \gm N^m 
\left ( 1 \; - \; \frac{N}{K} \right )^n \;  .
\ee
The solution to this equation could be reasonably well fitted to the world
population dynamics. The form of the solution is a slightly modified sigmoid 
curve, with the population never surmounting the carrying capacity $K$.  
For $m=1$ and $n=1$,  the Verhulst logistic equation (\ref{6}) is recovered.

\vskip 2mm

(x) {\bf Hyperlogistic time-delay equations}

\vskip 2mm

The straightforward extension of the Peschel-Mende hyperlogistic equation 
is the hyperlogistic time-delay equation
\be
\label{8b}  
\frac{dN(t)}{dt} = \gm N^m(t) 
\left [ 1 \; - \; \frac{N(t-\tau)}{K} \right ]^n \;  ,
\ee
which can also be treated as a generalization of the Hutchinson delayed 
logistic equation ({\ref{7}). This time-delay equation is capable of 
simulating a population that can essentially surmount the carrying 
capacity $K$ for a limited period of time, then dropping below it 
subsequently \cite{19}.     

\vskip 2mm

(xi) {\bf Singular Malthus equations}

\vskip 2mm

All equations enumerated above produce bounded solutions. In some cases,
the population dynamics seems to follow a law ending with divergent 
solutions. The well known Malthus equation \cite{20} gives the exponential 
population growth. But sometimes, the population dynamics develops a
super-exponential behavior, diverging at a finite time $t_c$ according to a power law
of the time to the singular time $t_c$.
The simplest way of modeling such a behavior is by the equation
\be
\label{8c}
 \frac{dN}{dt} = \gm N^m \;  ,
\ee
with the power $m \geq 1$. This is a direct generalization of the Malthus 
equation, capturing the positive feedback of the population on the growth rate: 
the larger the population, the higher the growth rate. For $m=1$, one recovers
the usual Malthus equation with the exponential solution. When $m > 1$, the 
solution to the equation (\ref{8c}) is of power law
$$
 N(t) = \frac{C}{(t_c-t)^{1/\ep}} \;  ,
$$
where
$$
C \; \equiv \; \frac{1}{(\ep\gm)^{1/\ep}} \; , \qquad 
\ep \equiv m - 1 \; ,
$$
$$
 t_c \; \equiv \; \frac{1}{N^\ep_0\ep\gm} \; , \qquad
N_0 \equiv N(0) \;  .
$$
The solution diverges hyperbolically at the critical time $t_c$. Such 
strongly singular solutions were first discussed by von Foerster 
et al. \cite{21} and applied to rationalize the super-exponential growth 
of the human world population \cite{22,23,24,25}, population dynamics and 
financial markets \cite{26,27,28}, material failures and 
earthquakes \cite{29,30}, climate and environmental changes \cite{31,32,33,34}, 
and dynamics of other systems \cite{35,36,37}. In ecology, the correlation 
between population density and the per capita growth rate is known as the
Allee effect \cite{38}. The feedback between the population density, 
associated with the Allee effect, can lead to the increase of the effective 
growth rate, in the case of sufficiently large populations, or to the rate 
decrease and species extinction for small-density populations \cite{38,39,40,41}. 

In addition to the differential equations describing population 
dynamics, there exist as well difference equations \cite{42,43} and 
integro-differential equations \cite{44,45}. There are more complicated
equations characterizing several factors, such as the population density, 
mass or weight dependence of individual members of different species, 
the dynamics of available food, and so on \cite{44,46,47}. Some study the influence 
on population dynamics of available information \cite{48}. It is also possible
to investigate the spatial dependence of populations \cite{49}. More details
on these and other types of equations can be found in the review articles 
\cite{50,51,52,53}.

We do not consider here the complications caused by the desire to take into
account many various features of the studied populations. 
Our aim here is different: we concentrate our attention on the influence of the
functional dependence of carrying capacities on population densities. Since 
this idea is rather new, it is necessary, first of all, to study the related 
effects for simpler equations, without overloading them by secondary 
specifications. Once the main influence of the functional carrying capacities 
is understood, it will be possible to complicate the equations by taking into 
account more and more mechanisms and specificities. For the same reason, 
the parameters characterizing the interactive species are treated as fixed.
In reality, the characteristics of each given biological species vary in general with
the age of the individuals composing the group. However, it is always admissible to 
divide the populations into different age ranges characterized by similar 
birth and death rates. Another possibility is to consider each species being
characterized by effective averaged parameters, which corresponds to what is
called the mean-field approximation \cite{36,37,54}.     

In the dynamics of any population, one can distinguish several time scales.
The shortest time scale is the {\it interaction time} $t_{int}$, describing 
interactions between individuals. This time is much shorter than the 
{\it observation time} $t_{obs}$, during which the population is investigated.
In order that the studied species could be characterized by fixed parameters,
such as birth-death rates, the observation time should be shorter than the
{\it variation time} $t_{var}$ during which individuals experience noticeable 
changes related, e.g., to their age. In this way, we keep in mind the situation,
when
$$
t_{int} \ll t_{obs} \ll t_{var} \; .
$$
Then all population parameters, except the carrying capacity, can be kept fixed.
Our main point is that the carrying capacity can vary due to mutual interactions 
of individuals, hence it varies during the interaction time and this variation 
needs to be taken into account.    

The standard situation in treating the population evolution by
means of the equations of the above types is that the carrying
capacities are kept as {\it fixed constant parameters}. In the
following Section 2, we propose an approach where the carrying
capacities are functionals of the species populations. This
makes it possible to drastically extend the applicability of
the population-evolution equations to various situations
describing the regimes that where unavailable with other models.
The basic idea of the approach has been formulated in our
previous papers \cite{39,40}, where some particular models were
considered. Now, in Section 2, we propose a general framework
for generating a large class of such models. In Sections 3 and 4,
we consider particular variants of the suggested equations,
corresponding to those of Refs. \cite{39,40}. The difference from
the previous works is three-fold. First, we simplify the
consideration by a convenient choice of the scaling for the
terms describing the carrying capacities, which allows us to make 
a more straightforward classification of different dynamical regimes.
Second, we emphasize the conditions under which solutions arise that 
are characterized by extreme events, such as the finite-time death 
and finite-time singularity of the species. Third, we suggest a new 
interpretation for an extreme event such as the finite-time 
singularity. The novel interpretation is based on the leverage effect 
and considers the singularity as a manifestation of an evolutional 
boom followed by a crash. Our considerations are phrased for 
applications to the development, growth and possible collapse of 
biological as well as human societies, as they both follow similar 
dynamics with analogous underlying mechanisms \cite{37}.

\section{Population evolution with functional carrying capacity}

The idea that the carrying capacity may be not a constant but a function
of population fractions has repeatedly appeared in the literature in the 
form of general discussions. In Sec. 2.1, we give a historical overview
of these ideas that provide a firm justification for their mathematical
representation in Sec. 2.2 and in the following sections.

\subsection{General meaning of functional carrying capacity}

The carrying capacity of a biological species in an environment is generally 
understood as the maximum population size of the species that the environment 
can sustain indefinitely, given the food, habitat, water and other necessities 
available in the environment. In population biology, carrying capacity is 
defined as the environment maximal load \cite{Hui2006}, which is different 
from the concept of population equilibrium. Historically, carrying capacity 
has been treated as a given fixed value \cite{Zimmerer1994, Sayre2008}. But
then, it has been understood that the carrying capacity of an environment may 
vary for different amounts of species and may change over time due to a variety 
of factors, including food availability, water supply, environmental conditions, 
living space, and population activity. 

Thus, the carrying capacity of a human society is influenced by the intensity 
of human activity, which depends on the level of technological development. 
When prehistoric humans first discovered that crude tools and weapons allowed 
greater effectiveness in gathering wild foods and hunting animals, they 
effectively increased the carrying capacity of the environment for their 
species. The subsequent development and improvement of agricultural systems 
has had a similar effect, as have discoveries in medicine and industrial 
technology. Clearly, the cultural and technological evolution of human 
socio-technological systems has allowed enormous increases to be achieved 
in carrying capacity for our species. This increased effectiveness of 
environmental exploitation has allowed a tremendous multiplication of the 
human population to occur \cite{Ricklefs1990, Freedman1995}.

Technology is an important factor in the dynamics of carrying capacity. For 
example, the Neolithic revolution increased the carrying capacity of the world 
relative to humans through the invention of agriculture. Currently, the use of 
fossil fuels has artificially increased the carrying capacity of the world by 
the use of stored sunlight, albeit with increasingly negative externalities,
such as global warming, ocean acidification and the indirect reduction of
diversity.  Other technological 
advances that have increased the carrying capacity of the world relative to 
humans are: polders, fertilizer, composting, greenhouses, land reclamation, 
and fish farming. Agricultural capability on Earth expanded in the last quarter 
of the 20th century. Whether this is sustainable is debatable.
There are signs that human-induced soil erosion as well as destabilization
of sensitive ecosystems may lead, at the same time, to a reduction of 
agricultural capability over the coming decades, such as in Africa
where the population is expected to double before 2050. The 
change in the carrying capacity of the habitat and environment 
supporting a human society can be described by the 
consumption impact \cite{Ehrlich1971}, which is proportional to the 
population size, with a coefficient characterizing the technology level.

One way to estimate human influence on the carrying capacity of the ecosystem is 
to use the so-called {\it ecological footprint accounting} method that provides empirical, non-speculative 
assessments of human activities with regard to the preservation or destruction of the 
Earth carrying capacity. It compares regeneration rates (biocapacity) against 
human demand (ecological footprint) in the same year. The results show that, in recent years,
humanity demand has exceeded the planet biocapacity by more than 20 
percent \cite{Wackernagel2002}.  
The present situation of rapid population growth in some regions, massive 
overexploitation of resources and steady accumulation of pollution and wastes 
diminishes the Earth carrying capacity. To a first approximation, with all
the caveats associated with the heterogeneity of technological developments
in different parts of the World, one can consider 
an average footprint per person, which leads to an 
estimation of the decrease of the 
Earth carrying capacity as roughly proportional to the size
of the total population. The question is how and by what means this
change of the Earth carrying capacity for humanity will pay back \cite{Dahl1996}.
 
Mutual coexistence and symbiosis of several species also strongly influences
the carrying capacities of the species, with the changes being, to a first
approximation, proportional to
the species numbers. For example, humans have increased the carrying capacity 
of the environment for a few other species, including those with which we live 
in a mutually beneficial symbiosis. Those companion species include more than 
about 20 billion domestic animals such as cows, horses, pigs, sheep, goats, 
dogs, cats, and chickens, as well as certain plants such as wheat, rice, barley, 
maize, tomato, and cabbage. Clearly, humans and their selected companions have 
benefited greatly through active management of mutual carrying capacities
\cite{Begon1990}.

Interactions between two or more biological species are known to essentially 
influence the carrying capacity of each other, by either increasing it, 
when species derive a mutual benefit, or decreasing it, when their interactions
are antagonistic \cite{Boucher1982, Callaway1995, Stachowicz2001}. The same
applies to economic and financial interactions between firms, which also form a 
kind of symbiosis, where the interacting firms develop the carrying capacity
of each other also roughly proportionally to their sizes \cite{Press2006}.

Many authors (e.g., Del Monte-Luna et al. \cite{Luna2004}) stress that, due 
to the influence on the carrying capacity resulting from the existing populations, its
original definition implying a constant value has lost its meaning. As a 
consequence of the feedback loops of the population sizes,
the notion of carrying capacities has taken a broader sense.
Carrying capacity should be understood as a 
non-equilibrium relationship or function that depends on the population size and 
on the symbiotic 
relations between the interacting populations. It characterizes the growth 
or development of available resources at all hierarchical levels of biological 
integration, beginning with the populations, and shaped by processes and 
interdependent relationships between finite resources and the consumers of 
those resources \cite{Luna2004}.

The above discussion shows that, in general, the carrying capacity is not a fixed 
quantity, but it should be considered a function of population sizes. In the case 
of a single species, the carrying capacity can be created or destroyed by the 
species activity. The simplest assumption is to take the species impact 
as proportional to the species population size.
When species coexist, their carrying capacities are influenced by the species
mutual interactions, either facilitating the capacity development or damaging 
it. Being functions of the species populations, such nonequilibrium carrying 
capacities can be naturally represented as polynomials over the population 
numbers \cite{Richerson1998}.

\subsection{Mathematical formulation of basic equations}

Let the considered society consist of several species enumerated
by an index $i = 1,2, \ldots$. The number of members in a society
is denoted as $N_i = N_i(t)$,

The main idea of our approach is that the carrying capacity of
each species is not a fixed constant, but it is a functional
\be
\label{9}
K_i = K_i( \{ N_i \} )
\ee
of the set $\{N_i\}$ of the species populations. This
assumption takes into account that the species may interact
not merely directly but also by influencing the carrying
capacities of each other as well as their own carrying
capacity.

A general form of the evolution equations, that takes into
account both direct interactions and mutual influence on the
carrying capacities, can be written as
\be
\label{10}
 \frac{dN_i}{dt} = \left ( \gm_i + \sum_{ij} 
\frac{C_{ij}}{K_i}\; N_j \right ) N_i \; ,
\ee
where $K_i$ is the functional carrying capacity (\ref{9}).
If the latter were a fixed parameter, one would return
to the generalized Lotka-Volterra predator-prey model (\ref{3}).
But in our approach, the carrying capacities are nontrivial
functionals of populations.

In a particular case, when the species do not display
strong direct interactions, but their mutual correlations
are mainly through influencing the carrying capacities of
each other, then Eq. (\ref{10}) reduces to the evolution
equation
\be
\label{11}
\frac{dN_i}{dt} = \gm_i N_i \; - \;
\frac{C_i N_i^2}{K_i} \; .
\ee
Here the effective rate
\be
\label{12}
\gm_i \equiv \gm_i^{birth} - \gm_i^{death}
\ee
is the difference between the birth and death rates of the
corresponding species. When birth prevails over death, then
$\gamma_i > 0$, while if death is prevailing over birth, then
$\gamma_i < 0$. In economic applications, birth translates
into gain and death into loss. The parameter
\be
\label{13}
C_i \equiv C_i^{comp} - C_i^{coop}
\ee
characterizes the difference between the competition and
cooperation of the members inside a given type of species. 
When competition is stronger than cooperation, then $C_i > 0$, 
while if cooperation is stronger, then $C_i < 0$.

It is important to stress that Eq. (\ref{10}) is principally
different, both mathematically as well as by its meaning,
from the predator-prey model (\ref{3}). Similarly,
Eq. (\ref{11}) is principally different from the logistic
equation (\ref{6}) or its variants (\ref{7}) and (\ref{8}).
As a result, the equations with functional carrying capacities
can display novel types of solutions allowing for the
consideration of effects that are absent in other evolution
equations. In the following sections, we consider concrete
examples of evolution equations with functional carrying
capacities.

\section{Action of society on its own carrying capacity}

\subsection{Justification for equation form}

In the previously studied variants of the logistic equation,
the carrying capacity is treated as a given quantity. However,
it is often the case that the society activity does influence
its own carrying capacity that can be either enhanced by producing
new goods, materials, knowledge, and so on, or can be destroyed by
unreasonable exploitation of resources, e.g., by deforestation,
polluting water, and spoiling climate. Therefore, the carrying
capacity, taking into account such feedback effects, must be
a functional $K = K(N)$ of the population $N$.

Thus, the population evolution is characterized by the equation
\be
\label{14}
\frac{dN}{dt} = \gm N \; - \; \frac{CN^2}{K(N)} \;  ,
\ee
with the carrying capacity depending on $N$ itself.

Moreover, the creation or destruction of the carrying capacity
by the society members does not occur immediately, but is delayed 
since any creation or destruction requires time for its realization. 
Hence, the variable $N$, entering the carrying capacity, should be 
delayed in time by a lag $\tau$, so that $K(N) = K(N(t-\tau))$.

Different types of carrying capacity can be introduced that depend 
on a delayed population variable. In the present paper, we consider 
a simple linear form
\be
\label{15}
 K(N) = A + B N( t - \tau)  \; .
\ee
Here the first term $A>0$ is a {\it natural carrying capacity},
provided by Nature. The second term is the created or
destroyed capacity, depending on whether the society activity
is constructive or destructive. The parameter $B$ is the
{\it production factor}, if it is positive, and it is a
{\it destruction factor}, when it is negative. Form (\ref{15})
of the carrying capacity agrees with the assumption of
{\it additivity}, when its different parts sum to produce the
total carrying capacity.

\subsection{Reduced quantities and choice of scaling}

As the population numbers can be very large, it is therefore more
convenient to deal with the reduced quantities
\be
\label{16}
x \equiv \frac{N(t)}{N_{eff} }
\ee
measured in units of some typical population size $N_{eff}$. It is 
also convenient to introduce dimensionless parameters for the 
natural carrying capacity
\be
\label{17}
a \equiv \frac{A}{N_{eff} } \left |\;
\frac{\gm}{C} \; \right |
\ee
and for the production-destruction factor
\be
\label{18}
b \equiv B \left |\; \frac{\gm}{C} \; \right | \;  .
\ee
The total dimensionless carrying capacity
\be
\label{19}
 y \equiv \frac{K(N)}{N_{eff} } \left |\;
\frac{\gm}{C} \; \right |
\ee
takes the form
\be
\label{20}
y = a + bx(t-\tau) \;   .
\ee

Up to now, the effective value $N_{eff}$ has been arbitrary.
By a special choice of the scaling, it is possible to simplify
the equations and to make a more transparent classification of
arising dynamical regimes. It is convenient to choose
\be
\label{21}
N_{eff} \equiv A \left |\; \frac{\gm}{C} \; \right | \; .
\ee
Then parameters (\ref{17}) and (\ref{18}) reduce to
\be
\label{22}
a = 1 \; , \qquad b = \frac{B}{A}\; N_{eff} \; .
\ee
The dimensionless carrying capacity (\ref{20}) becomes
\be
\label{23}
y = 1 + bx(t-\tau) \;   .
\ee

Let us also define the signs of the birth-death rate and of
the competition-cooperation parameter as
\be
\label{24}
\sgm_1 \equiv {\rm sgn} \gm = \frac{\gm}{|\gm|} \; , \qquad
\sgm_2 \equiv {\rm sgn} C = \frac{C}{|C|} \;   .
\ee
Depending on these signs, the following situations can occur.
\begin{eqnarray}
\label{25}
\begin{array}{lll}
\sgm_1 = +1 \; , & ~\sgm_2 = +1 & ~(gain \; + \; competition) \; , \\
\sgm_1 = +1 \; , & ~\sgm_2 = -1 & ~(gain\; + \;cooperation) \; , \\
\sgm_1 = -1 \; , & ~\sgm_2 = +1 & ~(loss\; + \;competition) \; , \\
\sgm_1 = -1 \; , & ~\sgm_2 = -1 & ~(loss\; + \;cooperation) \; .
\end{array}
\end{eqnarray}

Using the above notations and measuring time $t>0$ in units
of $1/\gamma$, we reduce  Eq. (\ref{14}) to
\be
\label{26}
\frac{dx}{dt} = \sgm_1 x - \sgm_2 \; \frac{x^2}{y} \;  ,
\ee
with the carrying capacity (\ref{23}). This equation is to be 
complemented by the initial conditions
$$
x(t) = x_0 \qquad (t\leq 0) \; ,
$$
\be
\label{27}
 y(t) = y_0 = 1 + bx_0 \qquad (t\leq 0) \; .
\ee
The solution for $x$, by its meaning, is to be positive.
The production-destruction factor $b$ can take any real values, being 
positive for the constructive society activity, while negative, for 
its destructive activity.

\subsection{Evolutionary stable states}

One of the most important problems in studying any
evolutional model is the determination of evolutionary
stable states. These are given by the stable stationary
solutions to the considered equation. In order to analyze the stability of the solutions 
to the differential delay equations, we employ the Lyapunov stability theory 
following the work by Pontryagin \cite{55} and the books \cite{11,15,16}.

The stationary states of Eq. (\ref{26}) are defined as the solutions to
the fixed-point equation
\be
\label{28}
\sgm_1 x^* \; - \; \frac{\sgm_2(x^*)^2}{1+ bx^*} = 0 \;  .
\ee
This yields two fixed points
\be
\label{29}
x_1^* = 0 \; , \qquad x_2^* = \frac{\sgm_1}{\sgm_2-b} \;  .
\ee

Resorting to the Lyapunov stability analysis, we need
to consider a small deviation
\be
\label{30}
\dlt x_j(t) = x_j(t) - x_j^* \qquad
(j=1,2)
\ee
from the related fixed point. This deviation satisfies the
equation
\be
\label{31}
 \frac{d}{dt}\; \dlt x_j(t) = C_j \dlt x_j(t)
+ D_j\dlt x_j(t-\tau) \; ,
\ee
in which
$$
C_j \equiv \sgm_1 \; - \;
\frac{2\sgm_2 x_j^*}{1+bx_j^*} \; , \qquad
D_j \equiv b\sgm_2 \frac{x_j^*}{1+bx_j^*} \;  .
$$
For the corresponding fixed points, these parameters are
$$
C_1 =\sgm_1 \; , \qquad D_1 = 0\; ,
$$
$$
C_2 = \sgm_1 \; \frac{b(\sgm_1-1)-\sgm_2}{b(\sgm_1-1)+\sgm_2} \; ,
\qquad
D_2 = \frac{b\sgm_2}{[b(\sgm_1-1)+\sgm_2]^2 } \; .
$$

Looking for the deviation in the exponential form
$$
\dlt x_j(t) \; \propto \; e^{\lbd_j t} \;   ,
$$
we obtain the equation
\be
\label{32}
\lbd_j = C_j + D_j e^{-\lbd_j\tau}
\ee
for the characteristic exponents $\lambda_j$. By using
the notation
$$
 W_j \equiv (\lbd_j-C_j) \tau \; , \qquad
z_j \equiv \tau D_j e^{-C_j\tau} \;  ,
$$
equation (\ref{32}) becomes 
$$
 W_j e^{W_j} = z_j \;  ,
$$
which is nothing but the equation defining the Lambert function $W_j$.
Therefore, the characteristic-exponent equation (\ref{32})
acquires the form
\be
\label{33}
\lbd_j = C_j + {1 \over \tau} W_j(D_j e^{-C_j\tau}) \;   .
\ee
The stationary solution is stable when the real part of the
characteristic exponent is negative, $\Re \lambda_j < 0$.

To proceed further, we shall analyze separately the cases
listed in Eq. (\ref{25}).

\section{Society with gain and competition}

Under the prevailing gain (birth) and competition, when
\be
\label{34}
\sgm_1 = 1 \; , \qquad \sgm_2 = 1 \;   ,
\ee
the evolution equation (\ref{26}) reads as
\be
\label{35}
 \frac{dx(t)}{dt} = x(t) \; - \;
\frac{x^2(t)}{1+bx(t-\tau)} \; .
\ee
At the initial stage, when $0 \leq t < \tau$, we have the
exact solution
$$
x(t) = \frac{x_0(1+bx_0)e^t}{1+ x_0(b-1+e^t)} \qquad
(0\leq t< \tau )  \;  .
$$
This can be used for constructing by iteration an approximate
solution at the second stage, when $\tau < t < 2\tau$. Then
we could find an approximate solution at the third step, and
so on. However, the accuracy of such iterative constructions 
quickly deteriorates and is admissible only over a couple of 
initial steps. More accurate solutions are to be found by numerically
solving the evolution equation (\ref{35}).

The stability analysis of the previous section shows that the
fixed point $x_1^* = 0$ is unstable for all $b$ and $\tau$.
The second fixed point
\be
\label{36}
x_2^* =\frac{1}{1-b} \equiv x^*
\ee
is stable when either
\be
\label{37}
 -1 < b < 1 \; , \qquad \tau \geq 0 \;   ,
\ee
or when
\be
\label{38}
 b < -1 \; , \qquad \tau\leq\tau_0 \; ,
\ee
where
\be
\label{39}
 \tau_0 \equiv \frac{1}{\sqrt{b^2-1} }\;
\arccos\left ( \frac{1}{b} \right ) \; .
\ee
The stability region is shown in Fig. 1.

Varying the system parameters and initial conditions, we
can meet the following dynamic regimes.

\subsection{Punctuated unlimited growth}

For the parameters
\be
\label{40}
 b \geq 1 \; , \qquad \tau \geq 0 \qquad (x_0 > 0 )\;  ,
\ee
the population grows by steps, as shown in Fig. 2. The
growth continues to infinite times. This is a typical
example of the punctuated evolution caused by the fact
that the production factor $b$ is positive and sufficiently
large. Hence, the carrying capacity is produced by the 
population, with a delay $\tau$.

\subsection{Punctuated growth to stationary state}

The punctuated growth is not always unbounded, but it can
be bounded by the fixed point, provided the initial condition
$x_0$ is smaller than $x^*$ and the parameters are
\be
\label{41}
0 \leq b < 1 \; , \qquad \tau\geq 0 \qquad
\left ( x_0 <x^* \right )   .
\ee
This regime is presented in Fig. 3.

\subsection{Punctuated decay to stationary state}

When the parameters are the same as in Eq. (\ref{41}), but
the initial condition $x_0$ is larger than the fixed
point $x^*$,
\be
\label{42}
x_0 > x^* = \frac{1}{1-b} \; ,
\ee
then there appears the punctuated decay, as illustrated
in Fig. 4.

\subsection{Punctuated alternation to stationary state}

If the carrying capacity is destroyed by the population,
then there can occur a punctuated alternation to a stationary
state, when the parameters and the initial condition are
\be
\label{43}
-1 \leq b < 0 \; , \qquad \tau\geq 0 \qquad
\left ( x_0 < \frac{1}{|b|} \right ) \; .
\ee
This is depicted in Fig. 5.

\subsection{Oscillatory approach to stationary state}

For sufficiently large destruction factor, there arises a
regime of an oscillatory approach to a stationary state, as
is presented in Fig. 6. This happens under the parameters
and the initial condition being defined by the inequalities
\be
\label{44}
b < -1 \; , \qquad  \tau <\tau_0 \qquad
\left ( x_0 < \frac{1}{|b|} \right ) \;   ,
\ee
where the lag $\tau_0$ is defined in Eq. (\ref{39}).

\subsection{Sustained oscillations}

There exists a lag $\tau_1=\tau_1(b)$, such that, when
\be
\label{45}
b < -1 \; , \qquad \tau_0\leq \tau \leq \tau_1 \qquad
\left ( x_0 < \frac{1}{|b|} \right ) \; ,
\ee
then oscillations do not decay, but continue without attenuation,
as in Fig. 7. The lag $\tau_1(b)$ can be found only numerically.

\subsection{Punctuated alternation to finite-time death}

If the time lag surpasses the value $\tau_1=\tau_1(b)$, the alternating
solution exists only for a limited time. At the death time $t_d$,
given by the equation
\be
\label{46}
 1 + b x(t_d-\tau) = 0 \; ,
\ee
all population becomes extinct, as in Fig. 8. This happens for
the parameters
\be
\label{47}
b < -1 \; , \qquad \tau > \tau_1 \qquad
\left ( x_0 < \frac{1}{|b|} \right ) \;   .
\ee

\subsection{Growth to finite-time singularity}

In the case, where the activity of the population is destructive,
time lags are large, and the initial condition is also large, so that
\be
\label{48}
b < 0 \; , \qquad \tau > \tau_c \qquad
\left ( x_0 > \frac{1}{|b|} \right ) \;   ,
\ee
the population dynamics becomes dramatic, diverging at a finite
time, called the {\it critical time}. The divergence is hyperbolic,
according to the law
\be
\label{49}
x(t) \; \propto \; \frac{1}{t_c-t} \qquad
(t\ra t_c-0 ) \;  ,
\ee
as is demonstrated in Fig. 9. The values of the critical lag
$\tau_c$ and the critical divergence time $t_c$ can be found
numerically.

\subsection{Unlimited exponential growth}

For shorter time lags, when
\be
\label{50}
 b < 0 \; , \qquad \tau \leq \tau_c \qquad
\left ( x_0 > \frac{1}{|b|} \right ) \;  ,
\ee
the divergence moves to infinity, the solution being a simple
growing exponential, as shown in Fig. 10.

\vskip 2mm

In this system with gain and competition, there may happen two 
extreme events, the finite-time death at a death time $t_d$ and 
the finite-time singularity at a critical time $t_c$. These two 
extreme events occur under the condition of a destructive activity 
of the population. The finite-time death is caused by the 
destruction of all resources. The finite time singularity implies 
that close to this critical point, the dynamic regime has to be changed,
according to the accepted interpretation of such singularities
\cite{27,39,40}. Such a change of the dynamic regime is analogous to
the occurrence of critical phenomena in statistical systems
\cite{36,37,54,56}. An interpretation of the finite-time singularity, based 
on the leverage effect, will be given below.

\section{Society with gain and cooperation}

When gain prevails over loss, and cooperation over competition,
that is, when
\be
\label{51}
 \sgm_1 = 1 \; , \qquad \sgm_2 = -1 \;   ,
\ee
the evolution equation takes the form
\be
\label{52}
\frac{dx(t)}{dt} = x(t) + \frac{x^2(t)}{1+bx(t-\tau)} \;  .
\ee

There are no stable stationary solutions in that case.
Depending on the system parameters and initial conditions,
there can arise the following dynamic regimes.

\subsection{Growth to finite-time singularity}

When the population activity is productive, but the time lag
is long, so that
\be
\label{53}
 b > 0 \; , \qquad \tau> \tau_c \qquad (x_0 > 0 ) \;  ,
\ee
or if the activity is destructive, when
\be
\label{54}
b < 0 \; , \qquad \tau > 0 \qquad
\left ( x_0 < \frac{1}{|b|} \right ) \;   ,
\ee
then the solution diverges at a finite critical time $t_c$.
The behavior is the same as in Fig. 9.

\subsection{Unlimited exponential growth}

Productive activity, under cooperation and not too long
time lags, such that
\be
\label{55}
b > 0 \; \qquad 0 < \tau \leq \tau_c \qquad (x_0>0) \;  ,
\ee
result in an exponential growth, as in Fig. 10.

\subsection{Punctuated unlimited growth}

For the parameters
\be
\label{56}
b < -1 \; , \qquad \tau\geq 0 \qquad
\left ( x_0 > \frac{1}{|b|-1} \right ) \;   ,
\ee
the solution displays unlimited punctuated growth, as in Fig. 2.

\subsection{Punctuated decay to finite-time death}

Destructive activity, under one of the conditions, when either
\be
\label{57}
 -1 < b < 0 \; , \qquad \tau \geq 0 \qquad
\left ( x_0 > \frac{1}{1-|b|} \right ) \;  ,
\ee
or when
\be
\label{58}
 b < -1 \; , \qquad \tau \geq 0 \qquad
\left ( x_0 < \frac{1}{|b|-1} \right ) \;  ,
\ee
leads to population extinction, at the death time given
by Eq. (\ref{46}). But the dynamics for this case, as is shown
in Fig 11, is different from that of Fig. 8. In the present case, 
there are no alternations, but the decay to zero is monotonic, 
exhibiting a finite number of quasi-plateaus.

\vskip 2mm

Under the conditions of gain and cooperation, there are two types 
of extreme events, the finite-time singularity at a critical 
time $t_c$ and the finite-time death at a death time $t_d$. The 
finite-time death is caused by the destructive population activity. 
And the finite-time singularity means that, close to the singularity 
point, the system experiences a change of dynamic regime.

\section{Society with loss and competition}

Under prevailing loss and competition, when
\be
\label{59}
 \sgm_1 = -1 \; , \qquad \sgm_2 = 1 \;   ,
\ee
the evolution equation becomes
\be
\label{60}
 \frac{dx(t)}{dt} = - x(t) \; - \;
\frac{x^2(t)}{1+bx(t-\tau)} \; .
\ee

There are two stable fixed points. One is the trivial point
\be
\label{61}
x_1^* = 0 \;  ,
\ee
which is stable for all parameters
\be
\label{62}
 -\infty < b < \infty \; , \qquad \tau \geq 0 \;  .
\ee
Another stationary point
\be
\label{63}
x_2^* = - \frac{1}{1+b} \equiv x^*
\ee
is stable for the parameters
\be
\label{64}
 b < -1 \; , \qquad \tau < \tau_0 \; ,
\ee
where
\be
\label{65}
 \tau_0 \equiv \frac{1}{\sqrt{b^2-1} }\; \arccos
\left ( \frac{1}{|b|} \right ) \;  .
\ee
Thus, there is the bistability region shown in Fig. 12.
Solutions tend to one of the two stationary states, when
the initial conditions are in the basin of attraction of the
corresponding fixed point. The following regimes can arise.

\subsection{Monotonic decay to zero}

In a society with prevailing loss and competition, the
decay to zero, as in Fig. 13, seems to be a natural type of
behavior. This happens when either
\be
\label{66}
 b > 0 \; , \qquad \tau \geq 0 \qquad
(x_0 > 0 ) \; ,
\ee
or when
\be
\label{67}
 b > 0 \; , \qquad \tau \geq 0 \qquad
\left ( x_0 < \frac{1}{|b|} \right ) \;  .
\ee

\subsection{Oscillatory convergence to stationary state}

When the parameters are such that
\be
\label{67a}
b < -1 \; ,  \qquad 0 < \tau < \tau_0 \qquad  
\left ( x_0 > \frac{1}{|b|-1} \right ) \; ,
\ee
the population fraction $x$ oscillates in time, converging
to the stationary state (\ref{63}), as is shown in Fig. 14.
Oscillations are caused by the presence of the time delay.

\subsection{Everlasting nondecaying oscillations}

For the parameters
\be
\label{68}
b < -1 \; , \qquad \tau_0 \leq \tau < \tau_1 \qquad
\left ( x_0 > \frac{1}{|b|-1} \right ) \;   ,
\ee
the solution oscillates without decay, similarly to the
behavior in Fig. 7. The time lag $\tau_0$ is given by Eq.
(\ref{65}) and $\tau_1$ is defined numerically.

\subsection{Punctuated growth to finite-time singularity}

A rather interesting behavior of the population dynamics
happens for the parameters
\be
\label{69}
 b < -1 \; , \qquad \tau_1 \leq \tau < \tau_2 \qquad
\left ( x_0 > \frac{1}{|b|-1} \right ) \;  .
\ee
Then the solution experiences several punctuations, after
which it diverges, as is illustrated in Fig. 15, at the
critical time defined by the equation
\be
\label{70}
 1 + bx(t_c-\tau) = 0 \;  .
\ee
When the final rise is preceded by a fall, this
behavior is reminiscent of the Parrondo effect \cite{57},

\subsection{Up-down convergence to stationary state}

A highly non-monotonic behavior exists for the parameters
\be
\label{71}
 b < -1 \; , \qquad 0 \leq \tau < \tau_c \qquad
\left ( \frac{1}{|b|} <  x_0 < \frac{1}{|b|-1} \right ) \;,
\ee
where the time lag $\tau_c$ can be found only numerically.
In this case, the solution, first, bursts out upwards, after which
it decays to the stationary value $x^*$, as in Fig. 16.

\subsection{Growth to finite-time singularity}

Under the parameters
\be
\label{72}
 b < -1 \; , \qquad \tau > \tau_c \qquad
\left ( \frac{1}{|b|} <  x_0 < \frac{1}{|b|-1} \right ) \; ,
\ee
the solution diverges at a finite critical time, without
any punctuation, in the same way as in Fig. 9.

\subsection{Unlimited exponential growth}

In the region of the parameters
\be
\label{73}
 -1 < b < 0 \; , \qquad 0 < \tau \leq \tau_c \qquad
\left ( x_0 > \frac{1}{|b|} \right ) \;  ,
\ee
the solution grows exponentially, as in Fig. 10.

\vskip 2mm

For a society with prevailing loss and competition, there
are two extreme events, both characterized by a finite-time
singularity at a critical time $t_c$. These regimes occur
under a strong destructive activity of the population and
a rather long time lag. The divergence can be understood
as a critical point where the society dynamics qualitatively 
changes.

\section{Society with loss and cooperation}

When loss and cooperation prevail, so that
\be
\label{74}
 \sgm = -1 \; , \qquad \sgm_2 = -1 \;  ,
\ee
the population evolution equation is
\be
\label{75}
 \frac{dx(t)}{dt} = - x(t) + \frac{x^2(t)}{1+bx(t-\tau)} \; .
\ee
There exists the sole evolutionary stable state
\be
\label{76}
 x^* = 0
\ee
that is stable for all parameters
\be
\label{77}
 -\infty < b < \infty \; , \qquad \tau \geq 0 \; .
\ee
The following dynamic regimes are possible.

\subsection{Monotonic decay to zero}

For the initial conditions in the attraction basin of the
stable fixed point, the solutions decay to zero with time,
as in Fig. 13. This happens when either
\be
\label{78}
 b < 0 \; , \qquad \tau \geq 0 \qquad
\left ( x_0 < \frac{1}{1-b} \right ) \;  ,
\ee
or when
\be
\label{79}
 b > 1 \; , \qquad \tau \geq 0 \qquad
\left ( x_0 > \frac{1}{b-1} \right ) \;  .
\ee

\subsection{Growth to finite-time singularity}

If the initial conditions are outside of the attraction
basin of the fixed point (\ref{76}), they can diverge
at a finite critical time, similarly to the behavior
in Fig. 9. This happens when either
\be
\label{80}
 b \leq 0 \; , \qquad \tau > 0 \qquad
\left ( x_0 < \frac{1}{|b|} \right ) \;  ,
\ee
or when
\be
\label{81}
0 < b < 1 \; , \qquad \tau > \tau_c \qquad
\left ( x_0 > \frac{1}{1-b} \right ) \;   .
\ee

\subsection{Unlimited exponential growth}

For the parameters
\be
\label{82}
 0 < b < 1 \; , \qquad \tau < \tau_c \qquad
\left ( x_0 > \frac{1}{1-b} \right ) \;  ,
\ee
the solution exhibits exponential growth, as in Fig. 10.

\subsection{Monotonic decay to finite-time death}

Finally, for the parameters
\be
\label{83}
 b < 0 \; , \qquad 0 \leq \tau < \tau_d \qquad
\left ( x_0 > \frac{1}{|b|} \right ) \;   ,
\ee
the population becomes extinct at a finite death time, as
in Fig. 17. The death time is defined by an equation having
the same form as Eq. (\ref{46}). However the decay to death
now is monotonic, which distinguishes it from the punctuated
behavior before death, shown in Fig. 8 and Fig. 11.

\vskip 2mm

The society with prevailing loss and cooperation can exhibit
the finite-time singularity as well as the finite-time death.
These two types of extreme events happen under the destructive 
activity of population.

Summarizing, all extreme events, except one, occur when the
population destroys its carrying capacity. The sole
exception is the case of a society with gain and cooperation,
when there can arise a finite-time singularity under $b>0$,
i.e., when the activity of the population is productive. This
latter type of finite-time singularity is analogous to
that studied in Ref. \cite{27}. Its appearance means that, near
the critical time, the society becomes unstable and requires
to change its parameters, for instance replacing cooperation
by competition. It seems to be rather clear that, when the
population grows too much, the competition of individuals must
come into play, becoming prevailing over their cooperation.
The finite-time singularities, occurring under the destructive
society activity, imply the existence of some critical events,
whose detailed interpretation will be given in Sec. 11.

\section{Mutual influence of symbiotic species on their
carrying capacities}

\subsection{Classification of symbiosis types}

When the considered society is structured with
several species, it is necessary to characterize their
interactions. The standard way of doing this is by assuming
the equations of the predator-prey type (\ref{3}), with
direct interactions of species that eat each other. Such
equations, however, cannot describe indirect interactions,
when the species do not kill each other, but influence the
carrying capacities of each other. Therefore, the predator-prey
equations are suitable for describing the predator-prey
relations, but are not suitable for characterizing symbiotic
relations \cite{40}.

Examples of symbiosis are ubiquitous in biology and ecology
\cite{58,59,60,61}. It is also widespread in human societies. 
For example, one can treat as symbiotic the interrelations between 
firms and banks, between population and government, between culture
and language, between economics and arts, and between basic
science and applied research.

Considering purely symbiotic relations, we need equation (\ref{11}), 
with the carrying capacities being functionals of the species 
populations. The natural form of such carrying capacities for 
symbiotic species is
\be
\label{84}
K_i = A_i + B_i S_i ( \{ N_i \} ) \;  .
\ee
Here $A_i > 0$ is the natural carrying capacity, provided by
nature, for an $i$-th species. The coefficient $B_i$
characterizes the strength of influence of other species on
the carrying capacity of the $i$-th species. When $B_i$ is
positive, it can be called the production factor, while,
if it is negative, it is the destruction factor. The function
$S(\{N_i\})$ is a symbiotic function specifying the mutual
relations between symbiotic species. Since the sign has
already been attributed to the factor $B_i$, the symbiotic
function can be treated as non-negative. 

Depending on the kinds of symbiotic relations, that is, on
the signs of the factors $B_i$, there can occur different
variants of symbiosis. To illustrate this, let us analyze the
case of two symbiotic species for which there can exist the
following types of symbiosis.

\vskip 2mm

(i) {\bf Mutualism}, when both species are useful for each
other, developing their mutual carrying capacities:
\be
\label{85}
 B_1 > 0 \; , \qquad B_2 > 0 \quad (mutualism) \; .
\ee

\vskip 2mm

(ii) {\bf Parasitism}, when one of the species is harmful
for another, or both species are harmful for each other,
destroying the carrying capacities, which happens under
one of the pairs of inequalities below:
\begin{eqnarray}
\label{86}
\begin{array}{lll}
B_1 > 0 \; , & \qquad B_2 < 0 \; , & \\
B_1 < 0 \; , & \qquad B_2 > 0 \; , & \quad (parasitism) \\
B_1 < 0 \; , & \qquad B_2 < 0 ~. &  
\end{array}
\end{eqnarray}

\vskip 2mm

(iii) {\bf Commensalism}, when one of the species is useful
for another, while the latter is indifferent to the existence
of the first species, which corresponds to the validity of one
of the pairs of equations:
\begin{eqnarray}
\label{87}
\begin{array}{lll}
B_1 > 0 \; , & \qquad B_2 = 0 \; , & \\
B_1 = 0 \; , & \qquad B_2 > 0 & \quad (commensalism) \;   .
\end{array}
\end{eqnarray}

\subsection{Normalized species fractions}

We continue analyzing the symbiotic coexistence of two kinds
of species. As always, it is more convenient to work with
reduced quantities. So, we introduce the reduced fractions
\be
\label{88}
x \equiv \frac{N_1}{N_{eff} } \; , \qquad
z \equiv \frac{N_2}{Z_{eff} } \;   ,
\ee
whose normalization values $N_{eff}$ and $Z_{eff}$ will be
chosen later. We define the dimensionless carrying capacities
\be
\label{89}
 y_1 \equiv \frac{\gm_1 K_1}{C_1 N_{eff} } \; , \qquad
y_2 \equiv \frac{\gm_1 K_2}{C_2 Z_{eff} }
\ee
and the relative birth rate
\be
\label{90}
\al \equiv \frac{\gm_2}{\gm_1} \;  .
\ee

With these notations, the symbiotic equations (\ref{11}), in
the case of two types of species, reduce to
\be
\label{91}
\frac{dx}{dt} = x \; - \; \frac{x^2}{y_1} \; , \qquad
 \frac{dz}{dt} =\al z \; - \; \frac{z^2}{y_2} \;  ,
\ee
where time is measured in units of $1/\gamma_1$. By their
definition, the solutions $x$ and $y$ are non-negative.
The equations are complemented by the initial conditions
\be
\label{92}
 x(0) = x_0 \; , \qquad z(0) = z_0 \; .
\ee

For the following analysis, it is necessary to make concrete
the explicit forms of the carrying capacities (\ref{84}).

\section{Symbiosis with mutual interactions}

\subsection{Derivation of normalized equations}

The action of the species on the carrying capacities of
each other can be different, depending on whether,
influencing the carrying capacities, the species interact
or not. If the species, in the process of influencing their
carrying capacities, interact with each other, then the
carrying capacities (\ref{84}) can be represented in the
form
\be
\label{93}
K_1 = A_1 + B_1 N_1 N_2 \; , \qquad
K_2 = A_2 + B_2 N_2 N_1  \;  .
\ee
Generally, the populations $N_1$ and $N_2$ in these
carrying capacities could depend on the shifted time, when
one would have
$$
K_i = A_i + B_i N_i(t-\tau_i) N_j(t-\tau_j) \;  ,
$$
where $i \neq j$. However, we need, first, to understand the
influence of symbiosis without the time lag. Therefore, we
consider below the interactions without time delay.

Introducing the dimensionless natural carrying capacities
\be
\label{94}
a_1 \equiv \frac{\gm_1 A_1}{C_1 N_{eff} } \; , \qquad
a_2 \equiv \frac{\gm_1 A_2}{C_2 Z_{eff} } \;
\ee
and dimensionless symbiotic factors
\be
\label{95}
b \equiv \frac{\gm_1 B_1 Z_{eff}}{C_1} \; , \qquad
g \equiv \frac{\gm_1 B_2N_{eff} }{C_2}
\ee
translates Eqs. (\ref{93}) into the dimensionless expressions
\be
\label{96}
y_1 = a_1 + bxz \; , \qquad y_2 = a_2 + gxz \;  .
\ee

Since the scaling values $N_{eff}$ and $Z_{eff}$ are arbitrary,
it is reasonable to choose them so as to simplify the equations.
For this purpose, we set
\be
\label{97}
 N_{eff} \equiv \frac{\gm_1 A_1}{C_1} \; , \qquad
Z_{eff} \equiv \frac{\gm_1 A_2}{C_2} \; .
\ee
Then the natural carrying capacities (\ref{94}) become
\be
\label{98}
a_1 = a_2 = 1 \;   .
\ee
And the total carrying capacities (\ref{96}) read as
\be
\label{99}
y_1 = 1 + bxz \; , \qquad y_2 = 1 + gxz \;   .
\ee

As usual, we measure time in units of $1/\gamma_1$. The most 
interesting case in symbiosis is when the species influence 
each other throughout their lifetimes, and when these lifetimes 
are of comparable durations. If this were not the case, i.e., with 
very different lifetimes, the symbiotic relations could not be 
supported for a duration longer than the shortest lifespan, making 
symbiosis inefficient for the longer-lived species. Therefore, we 
assume that the symbiotic species have comparable growth rates, 
because the inverse of the growth rate sets the time scale of 
lifetime, and the later is often found proportional to the growth 
period, at least for mammals  \cite{62}. We thus set $\alpha = 1$ and 
obtain the equations
\be
\label{100}
 \frac{dx}{dt} = x \; - \; \frac{x^2}{1+bxz} \; , \qquad
\frac{dz}{dt} = z \; - \; \frac{z^2}{1+gxz} \; .
\ee
We can note that these equations are symmetric with respect to
the simultaneous interchange between $x$ with $z$ and between $b$ 
with $g$. This symmetry will result in the corresponding symmetry 
of the following solutions.

\subsection{Evolutionary stable states}

Again we use the Lyapunov stability analysis \cite{11,16,55}.
Equations (\ref{100}) possess the non-zero stationary state
$$
x^* = \frac{1}{2g} \left [ 1 - b + g -
\sqrt{(1+b-g)^2 - 4b} \right ] \; ,
$$
\be
\label{101}
z^* = \frac{1}{2b} \left [ 1 + b - g -
\sqrt{(1+b-g)^2 - 4b} \right ] \;   .
\ee
It is stable when either
\be
\label{102}
b< 0 \; , \qquad -\infty < g < +\infty \;  ,
\ee
or when
\be
\label{103}
 0 \leq b < 1 \; , \qquad g\leq g_c \; ,
\ee
or when
\be
\label{104}
b \geq 1 \; , \qquad g\leq 0 \;  ,
\ee
where the critical value $g_c$ is
\be
\label{105}
 g_c \equiv \left ( 1 - \sqrt{b}\right )^2 \leq 1 \; .
\ee
The stability region is depicted in Fig. 18. The basin of attraction
of this stationary state, depending on the signs of
the symbiotic factors, is defined by the following equations:
$$
x_0z_0 < \frac{1}{|b|} \qquad ( b < 0 , \; g > 0 ) \; ,
$$
$$
x_0z_0 < \frac{1}{|g|} \qquad ( b > 0 , \; g < 0 ) \; ,
$$
\be
\label{106}
x_0z_0 <  min \left \{ \frac{1}{|b|}\; , \frac{1}{|g|}
\right \} \qquad ( b < 0 , \; g<0) \;  .
\ee

The solutions to the symbiotic equations (\ref{100}) should
be compared to those of the uncoupled equations
\be
\label{107}
 \frac{dx}{dt} = x - x^2 \; , \qquad
\frac{dz}{dt} = z - z^2 \qquad ( b = g = 0 )  \; ,
\ee
corresponding to the case of no symbiosis, when the stationary
states are $x^* = z^* = 1$.

Solving numerically the system of equations (\ref{100}) for
different symbiotic factors and initial conditions yields
the following possible dynamic regimes.

\subsection{Convergence to stationary states}

For the system parameters in the region of stability and
for the initial conditions in the basin of attraction of the 
non-zero fixed point (\ref{101}), both species develop and converge 
to the stationary state. This is illustrated in Fig. 19 for 
different types of symbiosis, where, for comparison, the solutions 
for the case of no symbiosis are also presented. The four possible 
cases are illustrated in the four panels of Fig. 19, depending on 
the relative positions of $x(t)$ and $z(t)$ compared with the solution 
of the uncoupled equations (\ref{107}).

\subsection{Unlimited exponential growth}

When stationary solutions do not exist, so that either
\be
\label{108}
 0 < b < 1 \; , \qquad g > g_c \;  ,
\ee
or when
\be
\label{109}
 b > 1 \; , \qquad g > 0 \; ,
\ee
or when they exist, but the initial conditions are taken
outside of the attraction basin, then the populations of
both species grow exponentially with time.

\subsection{Finite-time death and singularity}

In the case of mutual parasitism, there can happen an extreme
solution when one of the species becomes extinct at a
finite critical time, while the other species displays a
finite-time singularity, as is shown in Fig. 20. This happens
when the initial conditions are outside of the attraction
basin so that either
\be
\label{110}
 \frac{1}{|b| } < x_0 z_0 <  \frac{1}{|g| } \qquad
( b < g < 0 ) \;  ,
\ee
or if
\be
\label{111}
\frac{1}{|g| } < x_0 z_0 <  \frac{1}{|b| } \qquad
( g < b < 0 ) \;   .
\ee
The critical time is defined by one of the corresponding
equations:
$$
 x(t_c) z(t_c) = \frac{1}{|b|} \qquad ( b < g < 0 ) \; ,
$$
\be
\label{112}
  x(t_c) z(t_c) = \frac{1}{|g|} \qquad ( g < b < 0 ) \;  .
\ee

The appearance of such an extreme solution is caused by the
mutual parasitism of the species, destroying the carrying
capacities of each other.

\section{Symbiosis without direct interactions}

\subsection{Derivation of symbiotic equations}

In many cases, symbiotic species influence each other by
increasing (improving) the carrying capacities of each other,
which does not involve direct interactions between the species. 
The most known example of this type is the symbiosis between
tree roots and fungi. In that case, the carrying capacities
(\ref{84}) can be written in the form
\be
\label{113}
K_1 = A_1 + B_1 N_2 \; , \qquad K_2 = A_2 + B_2 N_1 \;  .
\ee
The dimensionless carrying capacities (\ref{89}) now read as
\be
\label{114}
 y_1 = a_1 + bz \; , \qquad y_2 = a_2 + gx \; .
\ee
Employing the scaling of Eqs. (\ref{97}) gives normalization
(\ref{98}) and the carrying capacities (\ref{114}) become
\be
\label{115}
 y_1 = 1 + bz \; , \qquad y_2 = 1 + gx \; .
\ee
Thus, we come to the symbiotic equations in dimensionless form
\be
\label{116}
\frac{dx}{dt} = x \; -\; \frac{x^2}{1+bz} \; , \qquad
\frac{dz}{dt} = z \; -\; \frac{z^2}{1+gx} \;    .
\ee
There exists again the symmetry with respect to the simultaneous 
interchange between $x$ and $z$ and between $b$ and $g$.

\subsection{Evolutionary stable states}

Equations (\ref{116}) possess a non-zero stationary state
\be
\label{117}
 x^* = \frac{1+b}{1-bg} \; , \qquad
z^* = \frac{1+g}{1-bg} \;  ,
\ee
which is stable when either
\be
\label{118}
 -1 \leq b < 0 \; , \qquad g \geq -1 \;  ,
\ee
or when
\be
\label{119}
 b \geq 0 \; , \qquad 0 \leq g \leq g_c \;  ,
\ee
where
\be
\label{120}
 g_c \equiv \frac{1}{b} \;  .
\ee
The stability region is presented in Fig. 21.

If the symbiotic relations correspond to mutualism or
commensalism, then the attraction basin of the stationary
solution (\ref{117}) is the whole region of positive
initial conditions:
\be
\label{121}
x _0 > 0 \; , \qquad z_0 > 0 \qquad
(b \geq 0, \; 0 \leq g < g_c ) \;  .
\ee
But if at least one of the species is parasitic, then the
attraction basins are defined by one of the conditions,
depending on the signs of the symbiotic factors:
$$
x_0 < \frac{1}{|g|} \; , \qquad z_0 > 0 \qquad
( b>0 , \; g < 0) \; ,
$$
$$
x_0 > 0 \; , \qquad z_0 < \frac{1}{|b|} \qquad
( b < 0 , \; g > 0) \; ,
$$
\be
\label{122}
x_0 < \frac{1}{|g|} \; , \qquad  z_0 < \frac{1}{|b|} \qquad
( b < 0 , \; g < 0) \;   .
\ee

The following dynamic regimes are possible.

\subsection{Convergence to stationary states}

If initial conditions are in the attraction basin, then
both species converge to their stationary populations.
The convergence can be monotonic or not, depending on the
system parameters and initial conditions, as is demonstrated
in Fig. 22.

\subsection{Unlimited exponential growth}

For the parameters outside the stability region, such that
\be
\label{123}
 b > 0 \; , \qquad g > g_c \;  ,
\ee
there exists a solution with exponential growth in time for both
species.

\subsection{Finite-time divergence}

Extreme solutions appear when at least one of the species is
parasitic and initial conditions are outside of the attraction
basin. Thus, when either
\be
\label{124}
x_0 > \frac{1}{|g|} \; , \qquad z_0 > 0 \qquad
( b > 0 , \; g < 0 ) \;  ,
\ee
or when
\be
\label{125}
x_0 > 0 \; , \qquad z_0 > \frac{1}{|b|}  \qquad
( b < 0 , \; g > 0 ) \;   ,
\ee
then one of the species experiences a finite-time singularity
at a critical time $t_c$ that is defined numerically. In this
case, when approaching $t_c$, one of the following behaviors
arise:
$$
x(t) \ra x(t_c) < \infty \; , \qquad z(t) \ra \infty \; ,
$$
\be
\label{126}
 x(t) \ra  \infty \; , \qquad z(t) \ra z(t_c) <  \infty \;  ,
\ee
where the first line corresponds to conditions (\ref{124}),
while the second line, to conditions (\ref{125}). The typical
behavior of populations is shown in Fig. 23.

\subsection{Finite-time extinction}

Parasitic symbiotic relations may end with one of the species
being extinct and the other continuing its life without
symbiosis. When either
\be
\label{127}
 x_0 < \frac{1}{|g|} \; , \qquad z_0 > 0 \qquad
( b > 0 , \; g\leq -1) \;   ,
\ee
or when
\be
\label{128}
  x_0 < \frac{1}{|g|} \; , \qquad z_0 > \frac{1}{|b|} \qquad
( b < 0 , \; g < 0) \;  ,
\ee
then the species $z$ dies at a finite time $t_d$, defined by
the relation
\be
\label{129}
 x(t_d) = \frac{1}{|g|} \;  .
\ee
That is, the species $x$ kills the species $z$:
\be
\label{130}
 x(t) \ra x(t_d) \; , \qquad z(t) \ra 0 \qquad
(t\ra t_d) \;  .
\ee

The opposite situation, when the species $z$ kills the
species $x$ occurs if either
\be
\label{131}
 x_0 > 0 \; , \qquad z_0 < \frac{1}{|b|} \qquad
( b \leq -1 , \; g > 0 ) \; ,
\ee
or if
\be
\label{132}
 x_0 > \frac{1}{|g|} \; , \qquad z_0 < \frac{1}{|b|} \qquad
( b < 0 , \; g < 0 ) \; .
\ee
Then the species $x$ dies at a finite time given by the relation
\be
\label{133}
 z(t_d) = \frac{1}{|b|} \;  ,
\ee
so that
\be
\label{134}
 x(t) \ra 0 \; , \qquad z(t) \ra z(t_d) \qquad
(t\ra t_d) \;   .
\ee
The corresponding behavior is illustrated in Fig. 24.

\section{Interpretation of extreme events in population evolution}

\subsection{Types of extreme events}

In the population evolution, there may happen two types of extreme 
events, finite-time death and finite-time singularity. The origin for 
the occurrence of finite-time death is rather clear. This happens when
the carrying capacity of the species is destroyed, either by the species 
themselves or by the parasitic symbiosis of other species. The destroyed 
carrying capacity makes it impossible the long-term existence of the species that, 
thus, go towards extinction. 

Finite-time singularity can be due to two causes. One reason is the 
existence of cooperation between the members of species, as in Secs. 5.1 
and 7.2. This type of the finite-time singularity means that the society,
in which cooperation persists under fast increasing numbers of its 
members, becomes unstable and, to be stabilized, requires that cooperation 
be changed into competition. The necessity for such a change looks rather 
evident and is easily understandable. Really, in the presence of a strongly 
increasing population, the competition for the means of survival will become
unavoidable. 

A more elaborate mechanism operates in the case of the finite-time 
singularities occurring under competition, as in Secs. 4.8, 6.4, 6.6, 9.5, 
and 10.5. In all these cases, the singularities appear under the destruction 
of the carrying capacities either by the society itself or by a parasitic 
symbiotic species. It may seem quite strange that, while the carrying 
capacity is being destroyed, the population continues growing. To understand 
the origin of such a paradoxical effect and of these finite-time 
singularities, let us consider in turn the different types of finite-time 
singularities found in Secs. 4.8, 6.4, 6.6, 9.5, and 10.5.

\subsection{Boom and crash in society with gain and competition}

This corresponds to the case studied in Sec. 4.8, of a finite-time 
singularity occurring at a critical time $t_c$. The divergence is of the 
hyperbolic type (\ref{49}). This extreme event happens under the destructive 
action of the society on its own carrying capacity, when $b<0$. The parameters 
are such that, at the initial moment of time, the effective carrying capacity 
is negative,
\be
\label{135}
 y(0) = 1 - |b| x_0 < 0 \; .
\ee

How would it be possible to understand the existence of a negative carrying 
capacity? For some simple biological species, as ants or bees, the negative 
capacity would, probably, be impossible. Such species would not be able to 
live at all. However, for more complex societies such as human societies, 
the negative carrying capacity may have sense. For instance, humans do extract
non-renewable resources that become progressively exhausted forever, they 
destroy their habitat, poison rivers, pollute air, cut forests, and so on. 
At the same time, humans possess the ability of regenerating the habitat by 
cleaning rivers, or even oceans, and planting trees. Thus, humans may spoil 
their habitat to such an extent that it would require a hard work for its 
recovering. In that sense the effective carrying capacity can become negative 
for a while, implying the necessity of its recuperation in the positive 
domain in order to ensure the long-term survival of the human society 
\cite{63}.   

Even more transparent is the explanation for the existence of negative 
carrying capacity for financial and economic societies, when the variable 
$x$ represents not population, but capitalization. In these cases, negative 
capacity is nothing but the borrowed resources that have to be returned back 
to the lender. Due to this leverage effect resulting from borrowing, a firm 
can exhibit a fast development. But borrowing cannot last forever. If the 
firm, society, or country does not produce enough and is not able to pay 
debts, its creditors will lose trust and may require early reimbursement, or 
will simply refuse rolling over the debts, as occurred for Greece in May 2010
and Ireland in November 2010. This situation can be captured by assuming the 
existence of a maximum level of debt, beyond which the society or country 
becomes highly unstable due to feedbacks resulting from market forces. 
Actually, Reinhart and Rogoff \cite{64} have recently documented 
the existence of a strong link between levels of debt and countries' economic 
growth over the last two centuries: Countries with a gross public debt 
exceeding about 90\% of annual economic output tended to grow a lot more slowly
and to exhibit larger default risks.

Assuming the existence of a maximum debt level beyond which instabilities 
appear leads to the existence of a time $t_{crash}$ beyond which a crash or 
at least strong turbulence can occur. This is highly reminiscent of the scenario 
leading to the ``great recession'' that started in 2007 worldwide 
\cite{65}. The minimum {\it crash time} $t_{crash}$ is thus given 
by the condition that the debt, represented by the negative carrying capacity, 
reaches the value
\be
\label{136}
 y(t_{crash}) = 1 - |b| x(t_{crash}-\tau) < 0 \;  .
\ee
The crash happens before the critical divergence time $t_c$,
\be
\label{137}
 t_{crash} < t_c \;  ,
\ee
where the firm or country capitalization is still finite. In such a regime, 
the accelerated growth, fueled by borrowing, leads to a boom that is not 
supported by increasing productivity. This can therefore be called a bubble 
\cite{28}. As the bubble develops, it eventually reaches a threshold 
level beyond which it becomes unstable, and can therefore be followed by a 
crash at times between $t_{crash}$ and $t_c$.

\subsection{Boom and crash in society with loss and competition}

The same interpretation as above is applicable for the society with loss 
and competition, as in Secs. 6.4 and 6.6. There, the finite-time singularity
arises under a high level of destruction, when the destruction coefficient 
$b<-1$ and the initial carrying capacity is negative. The hyperbolic 
divergence occurs at a critical time $t_c$. In Sec. 6.6, the situation is 
similar to that discussed above. The difference between Sec. 6.4 and Sec. 6.6 
is that in Sec. 6.4 the divergence is defined by the equation
\be
\label{138}   
 y(t_c) = 1 - |b| x(t_c-\tau) = 0 \;  .
\ee

Again, a society or a firm with loss and competition, actually, does not 
reach the point of divergence, but becomes bankrupt before this. The fast 
growth is due to exploiting and destroying the carrying capacity. But, 
after destruction has taken place and reached an unbearable level, 
the boom is followed by a crash.

\subsection{Species extinction under mutual parasitic symbiosis}

In the parasitic symbiosis of two species considered in Sec. 9.5, 
there occurs a finite-time singularity. Thus, for the symbiotic 
parameters $b<g<0$, the initial carrying capacities are such that  
\be
\label{139}
 y_1(0) = 1 - |b| x_0 z_0 < 0 \; , \qquad
 y_2(0) = 1 - |g| x_0 z_0 > 0 \; .
\ee
For the opposite case, when $g<b<0$, the situation is symmetric. Thence,
below we shall treat the case of Eq. (\ref{139}) without loss of 
generality. The divergence appears at the critical time $t_c$ given by 
the equation
\be
\label{140}
 y_1(t_c) = 1 - |b| x(t_c) z(t_c)  \;  .
\ee
At this time, the population of the species $x$ tends to infinity, while 
that of the species $z$ goes to zero.

Of course, no realistic population can rise to infinite values. Such a 
divergence happens because of the mutual parasitic symbiotic relations,
resulting in the formal appearance of a negative effective capacity. 
As in the cases above, the divergence can be avoided by limiting the 
carrying capacity by a fixed level. This implies that the rise of a 
parasitic species continues only up to some limiting carrying capacity 
threshold
\be
\label{141}
 y_1(t_{crash}) = 1 - |b| x(t_{crash}) z(t_{crash}) \;  ,
\ee
after which the species $x$ dies out by a fast process of extinction at 
the crash time $t_{crash} < t_c$.

\subsection{Species extinction under parasitic symbiosis without direct 
interactions}

A finite-time singularity also appears in the case of symbiosis without
direct interactions, as in Sec. 10.5. This happens when at least one of the 
species is parasitic. For example, for the case $b<0$ and $b<g$. Below, we 
shall consider this case, since the situation with $g<0$ and $g<b$ is 
symmetric.

When $b<0$, this means that the species $z$ is parasitic and destroys the
carrying capacity of the species $x$. The finite-time singularity occurs
if, at the initial moment of time, the effective carrying capacity of species 
$x$ is negative,
\be
\label{142}  
 y_1(0) = 1 - |b| z_0 < 0 \;  ,
\ee
while the carrying capacity of species $z$,
\be
\label{143}
 y_2(0) = 1 + gx_0 \;  ,
\ee
can be positive or negative, depending on the values of $g$ and $x_0$. The 
divergence of $x$ occurs at the critical time $t_c$, where the effective 
capacity of species $x$ is negative,
\be
\label{144}
  y_1(t_c) = 1 - |b|z(t_c) < 0 \; .
\ee
The population of species $z$ at the moment $t_c$ is finite.

In the same way as in the previous cases, we understand that this 
divergence cannot be real and there should exist a limiting carrying 
capacity
\be
\label{145}
 y_1(t_{crash}) = 1 - |b| z(t_{crash}) \;  ,
\ee
at which the population $x$ is to be set to zero, implying its 
extinction caused by the parasitic species $z$. This extinction happens 
at the crash time $t_{crash} < t_c$.

\vskip 2mm

In all these cases for which there arises a finite-time singularity, it 
is possible to exclude the formal divergence by limiting the carrying 
capacity to a minimal value $y(t_{crash})$, that is, a maximal absolute 
value $|y(t_{crash})|$. This limiting value can be interpreted as a 
threshold for a change of regime. The overall dynamics, thus, starting 
with the fast growth of the population (or capitalization), is followed 
by its drop to zero at the crash time $t_{crash}$, before the critical 
time $t_c$.

\section{Conclusion}

In this paper, we have suggested a general approach for
describing the evolution of populations, whose activities
influence their carrying capacities. In order to take into
account this influence, the carrying capacities are to be
defined as functions of the society populations. This includes 
the action of a population on its own carrying capacity. 
In general, the actions of populations on the carrying 
capacities can be delayed, since such actions, generally, 
require time for their realization.

The approach is illustrated by analyzing the time evolution
of a society that acts on its own carrying capacity, either by
producing the increase of the capacity or by destroying it.
Different types of societies have been studied, depending on
the balance between gain and loss and between competition and
cooperation. A detailed classification of admissible dynamic
regimes has been given.

Two kinds of extreme events have been found to arise, when the 
society destroys its carrying capacity. One is a finite-time death
at a death time $t_d$ and another is a finite-time singularity
at a critical time $t_c$. The finite-time death describes
the extinction of the population because of the destruction
of the carrying capacity. The finite-time singularity signals
that the society becomes unstable and its stabilization
requires changing the society parameters and a transfer to
another dynamic regime. The divergence can be avoided by 
limiting the carrying capacity and interpreting the effect
as a fast rise of the population (or capitalization), followed 
by its sharp drop. For economic and financial societies, the 
fast growth is understood as a boom or bubble, due to the 
leverage effect induced by over-indebtedness, after which a crash occurs.  

The suggested approach is also illustrated by considering the
symbiosis of several species. This approach allows us to give
a general classification of different symbiosis types. The
case of two species is analyzed in detail. Extreme events arise
when at least one of the species is parasitic, destroying the
carrying capacity of the other species. Again, there can exist
two kinds of such extreme events, finite-time death and
finite-time singularity. Their interpretation is analogous to
that given for the case of the self-destructing population
activity.

As a general conclusion valid for the different considered
situations, we have to say that any destructive action of
populations, whether on their own carrying capacity or on the
carrying capacities of co-existing species, can lead to the
instability of the society that is revealed in the form of the 
appearance of extreme events, finite-time extinctions or booms 
followed by crashes.

\vskip 1cm
{\bf Acknowledgement}: We acknowledge financial support from the
ETH Competence Center ``Coping with Crises in Complex Socio-Economic
Systems" (CCSS) through ETH Research Grant CH1-01-08-2 and ETH Zurich
Foundation.

\newpage

\begin{figure}[ht]
\centerline{\includegraphics[width=10cm]{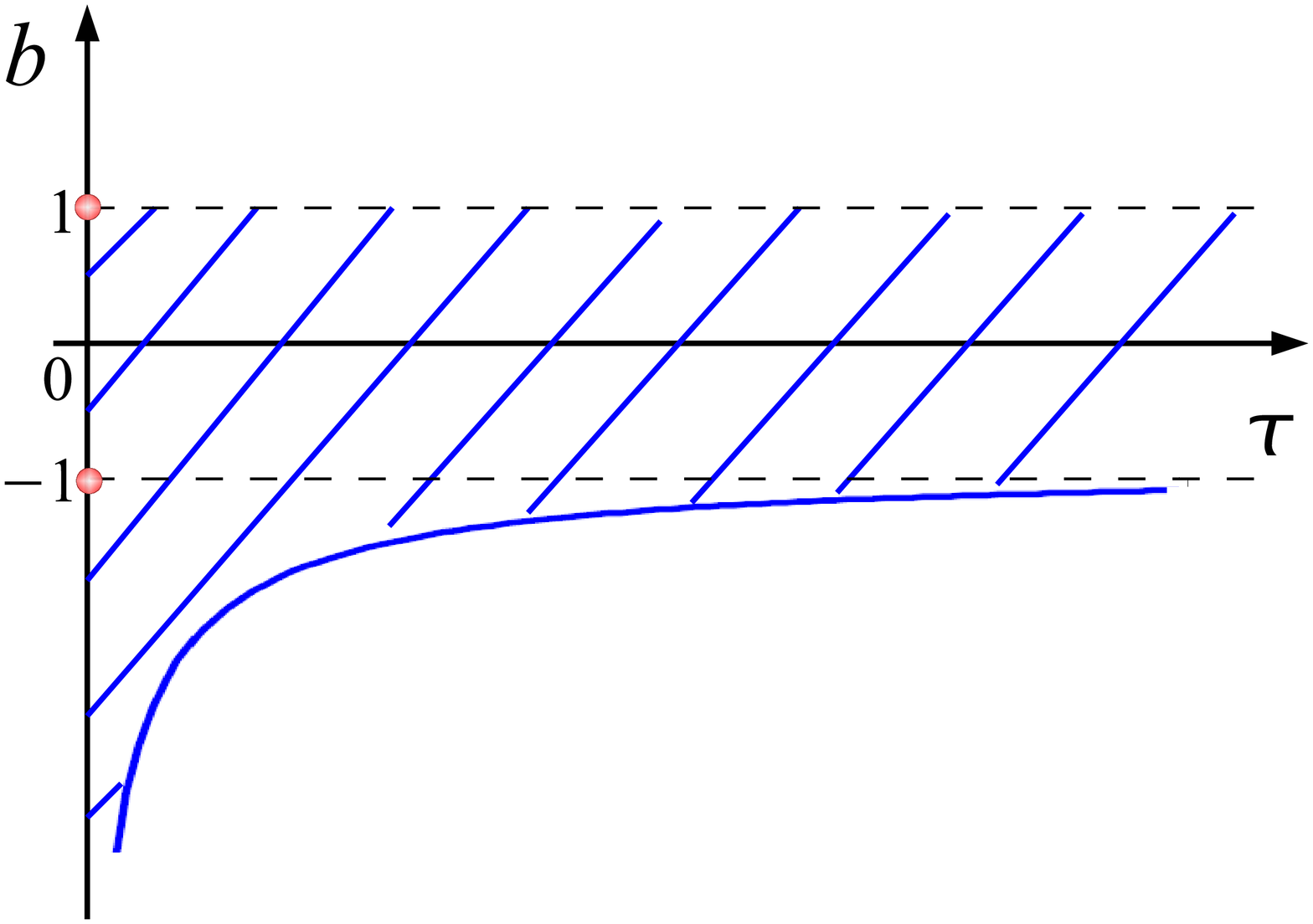}}
\caption{Stability region for the gain-competition case
defined in (\ref{34}).}
\label{fig:Fig.1}
\end{figure}

\newpage

\begin{figure}[ht]
\centerline{\includegraphics[width=10cm]{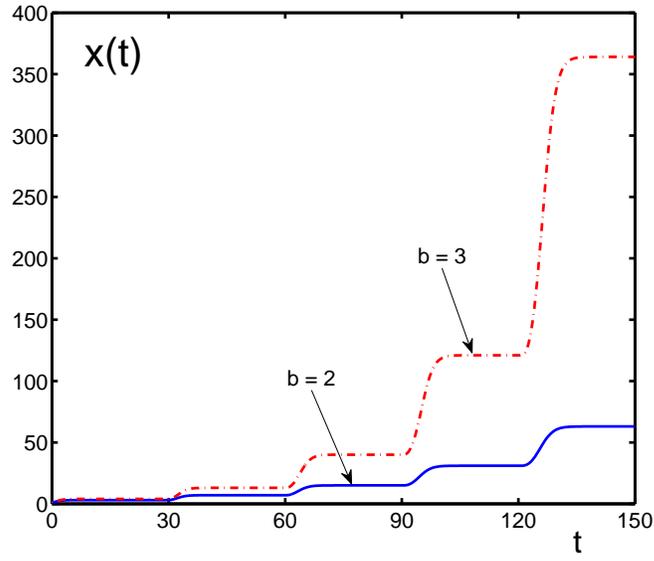}}
\caption{Temporal behavior of solutions to equation (\ref{35}) for
the initial condition $x_0=1$, lag $\tau=30$, and parameters
$b=2$ (solid line) and $b=3$ (dashed-dotted line). Solutions
$x(t)\ra\infty$, when $t\ra\infty$.}
\label{fig:Fig.2}
\end{figure}

\newpage

\begin{figure}[ht]
\centerline{\includegraphics[width=10cm]{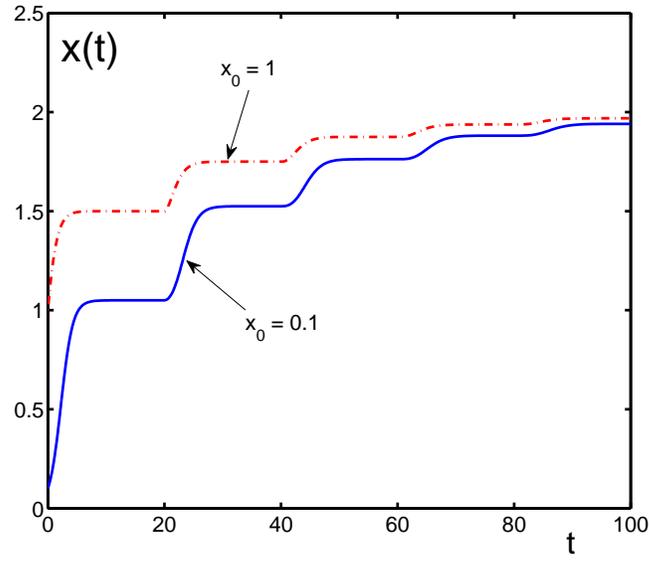}}
\caption{Dynamics of solutions to equation (\ref{35}) for
the lag $\tau=20$, symbiotic parameter $b=0.5$, and the 
initial conditions $x_0=0.1$, (solid line) and $x_0=1$ 
(dashed-dotted line). Solutions $x(t)\ra x^*$, monotonically 
growing by steps, when $t\ra\infty$. The stationary point 
is $x^*=1/(1-b)=2$.}
\label{fig:Fig.3}
\end{figure}

\newpage

\begin{figure}[ht]
\centerline{\includegraphics[width=10cm]{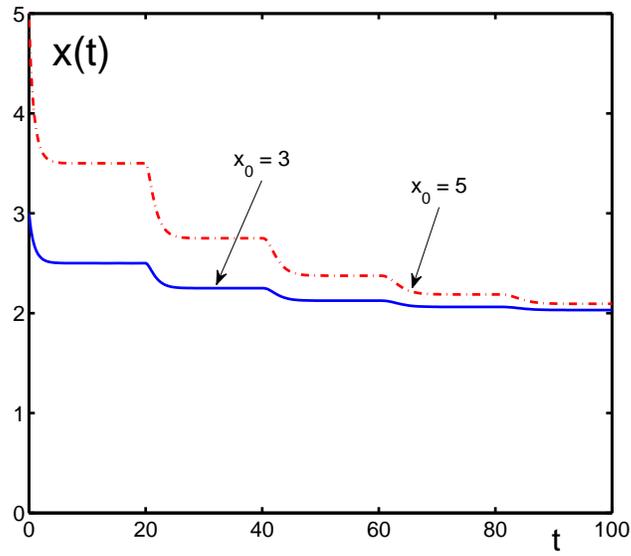}}
\caption{Evolution of solutions to equation (\ref{35}) for
the lag $\tau=20$, parameter $b=0.5$, and initial conditions
$x_0=3$, (solid line) and $x_0=5$ (dashed-dotted line).
Solutions $x(t)\ra x^*$, monotonically diminishing by steps,
when $t\ra\infty$. The stationary point is $x^*=1/(1-b)=2$, 
the same as in Fig. 3.}
\label{fig:Fig.4}
\end{figure}

\newpage

\begin{figure}[ht]
\centerline{\includegraphics[width=10cm]{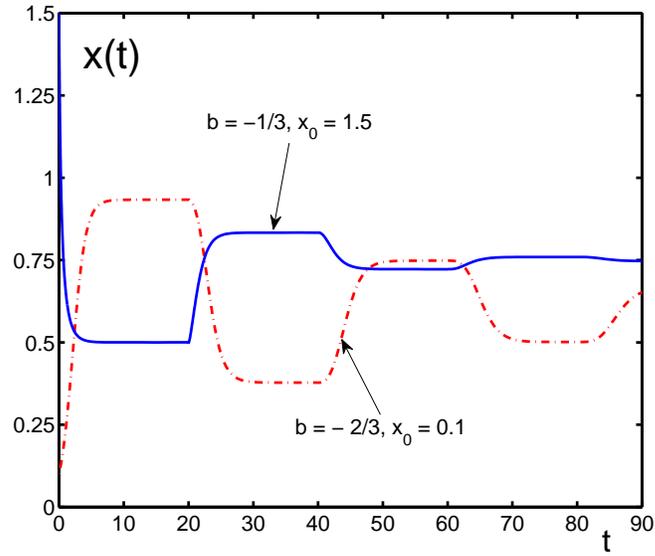}}
\caption{Temporal behavior of solutions to equation (\ref{35}) for
the lag $\tau=20$, parameters $b=-1/3$, $x_0=1.5$ (solid line),
and $b=-2/3$, $x_0=0.1$ (dashed-dotted line).
Solutions $x(t)\ra x^*$, oscillating by steps,
when $t\ra\infty$. The stationary points are $x^*=0.75$ and 
$x^*=0.6$, respectively.}
\label{fig:Fig.5}
\end{figure}

\newpage

\begin{figure}[ht]
\centerline{\includegraphics[width=10cm]{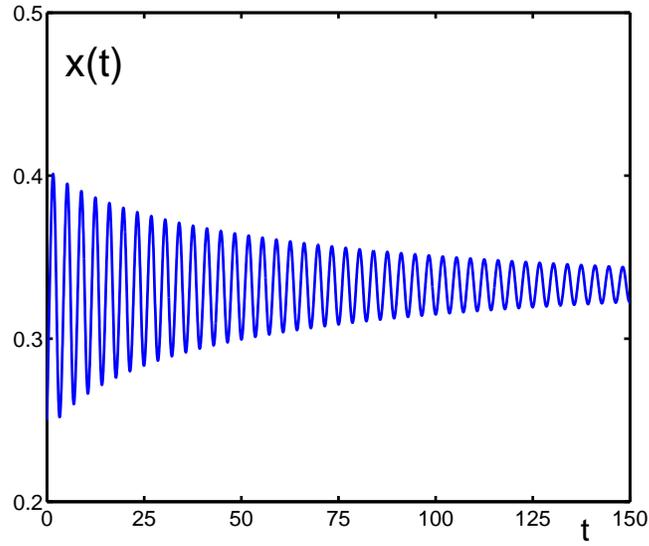}}
\caption{Temporal behavior of the solution to equation (\ref{35}) for the
parameter $b=-2<-1$, lag $\tau=1.18<\tau_0$, where $\tau_0=1.2092$ 
is defined by (\ref{39}), with the initial condition $x_0=0.25<1/|x_0|$.
Solution $x(t)\ra x^*=1/3$, oscillating, when $t\ra\infty$.}
\label{fig:Fig.6}
\end{figure}

\newpage

\begin{figure}[ht]
\centerline{\includegraphics[width=10cm]{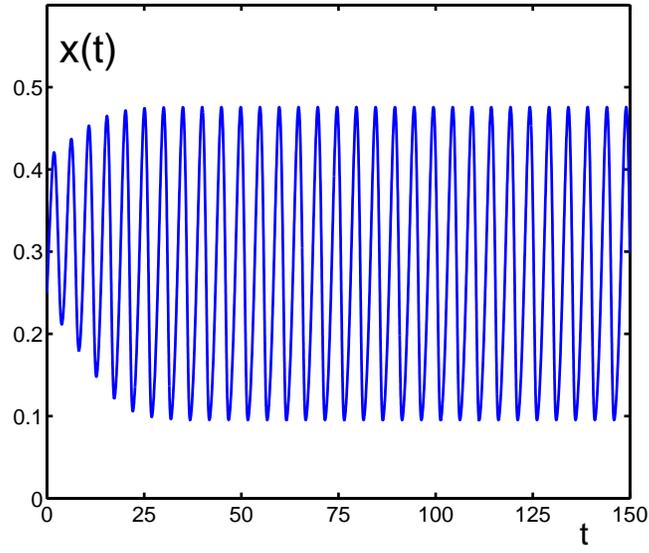}}
\caption{Dynamics of the solution to equation (\ref{35}) for
the parameter $b=-2<-1$, lag $\tau_0<\tau=1.5<\tau_1$, where
$\tau_0=1.2092$ is defined by (\ref{39}), $\tau_1=\tau_1(b)
\approx 1.65$, with the initial condition $x_0=0.25<1/|b|$. Solution
$x(t)$ oscillates without convergence, when $t\ra\infty$.}
\label{fig:Fig.7}
\end{figure}

\newpage

\begin{figure}[ht]
\centerline{\includegraphics[width=10cm]{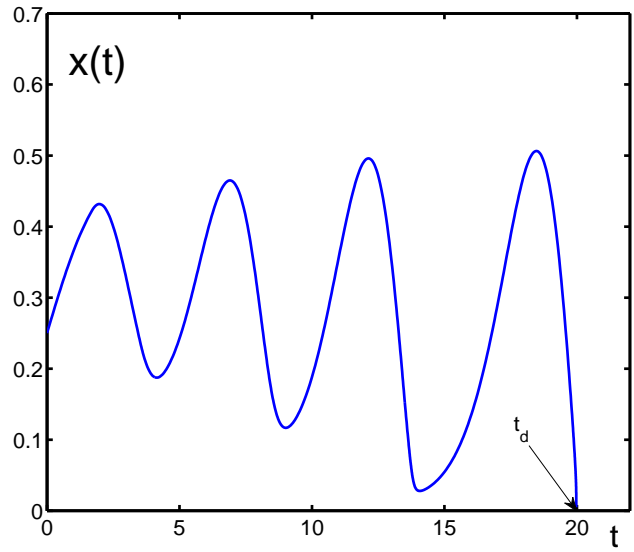}}
\caption{Dynamics of the solution to equation (\ref{35}) for
the parameter $b=-2<-1$, lag $\tau=1.7>\tau_1$, where $\tau_1
\approx 1.65$ is defined numerically, with the initial condition
$x_0=0.25<1/|b|$. Solution $x(t)$ exists only till the moment
$t=t_d\approx 19.975$, defined by (\ref{46}).}
\label{fig:Fig.8}
\end{figure}

\newpage

\begin{figure}[ht]
\centerline{\includegraphics[width=10cm]{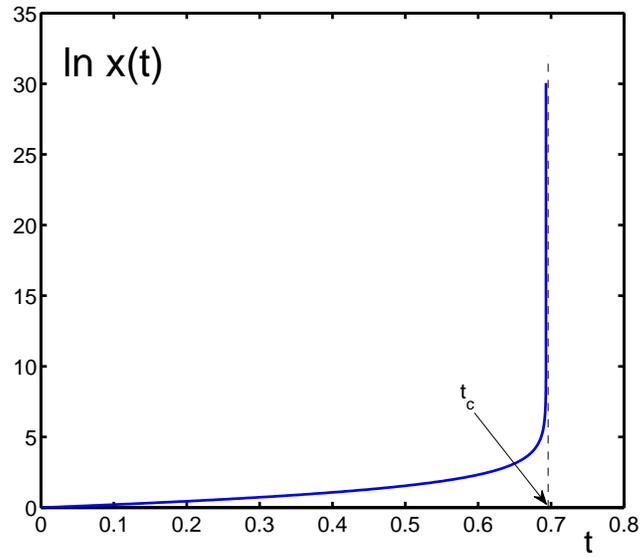}}
\caption{Behavior of the solution to equation
(\ref{35}) in logarithmic scale for the parameter 
$b=-2<-1$, lag $\tau=1.7>\tau_1$, where $\tau_1\approx 1.65$ is 
defined numerically, and the initial condition $x_0=1>1/|b|$. 
Solution $x(t)$ diverges, $x(t)\ra\infty$, when $t\ra t_c\approx 0.69315$.}
\label{fig:Fig.9}
\end{figure}

\newpage

\begin{figure}[ht]
\centerline{\includegraphics[width=10cm]{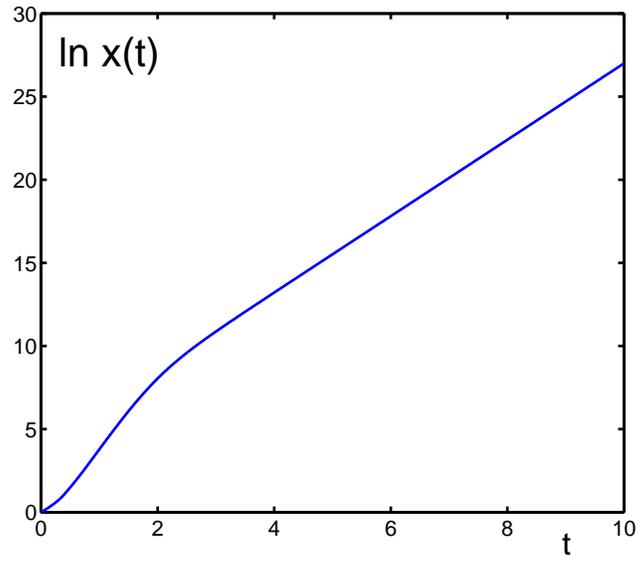}}
\caption{Behavior of the solution, shown in logarithmic scale, to equation
(\ref{35}) for the parameter $b=-2<-1$, lag $\tau=0.415<\tau_c$,
where $\tau_c\approx 0.416$ is defined numerically, with the initial
condition $x_0=1>1/|b|$. Solution $x(t)$ exponentially grows, when
$t\ra\infty$.}
\label{fig:Fig.10}
\end{figure}

\newpage

\begin{figure}[ht]
\centerline{\includegraphics[width=10cm]{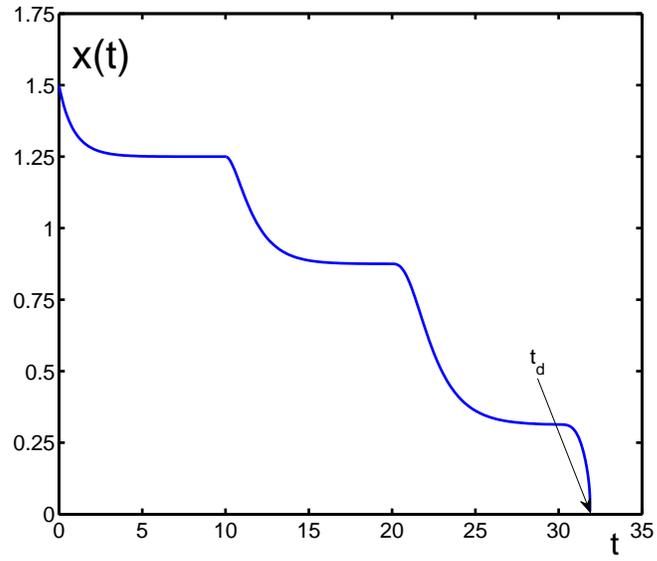}}
\caption{Evolution of the solution to equation (\ref{52}) for the 
parameter $b=-1.5<-1$, lag $\tau=10$, with the initial
condition $x_0=1.5<1/(|b|-1)$. Solution $x(t)$ monotonically decays
to finite-time death at $t_d\approx 31.884$, defined by (\ref{46}).}
\label{fig:Fig.11}
\end{figure}

\newpage

\begin{figure}[ht]
\centerline{\includegraphics[width=10cm]{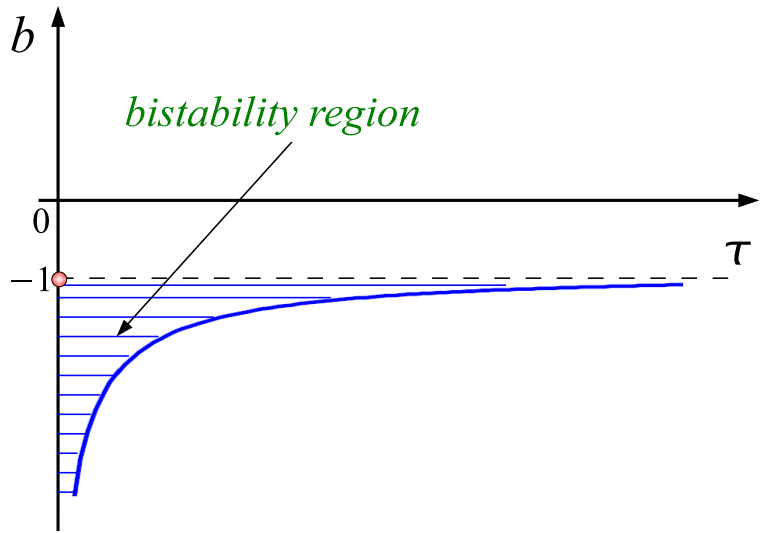}}
\caption{Stability region for the loss-competition case defined
in (\ref{59}).}
\label{fig:Fig.12}
\end{figure}

\newpage

\begin{figure}[ht]
\centerline{\includegraphics[width=10cm]{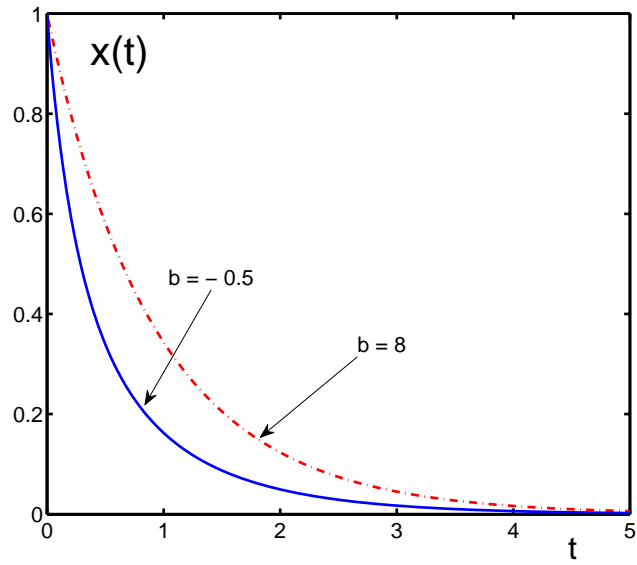}}
\caption{Temporal behavior of the solution to equation
(\ref{60}) for the lag $\tau=10$, initial condition $x_0=1$, with the 
parameters $b=-0.5$ (solid line) and $b=8$ (dashed-dotted line).
The initial condition $x_0<1/|b|$ for $b<0$. Solutions
$x(t)$ monotonically converge to their stationary state $x^*=0$.}
\label{fig:Fig.13}
\end{figure}

\newpage

\begin{figure}[ht]
\centerline{\includegraphics[width=10cm]{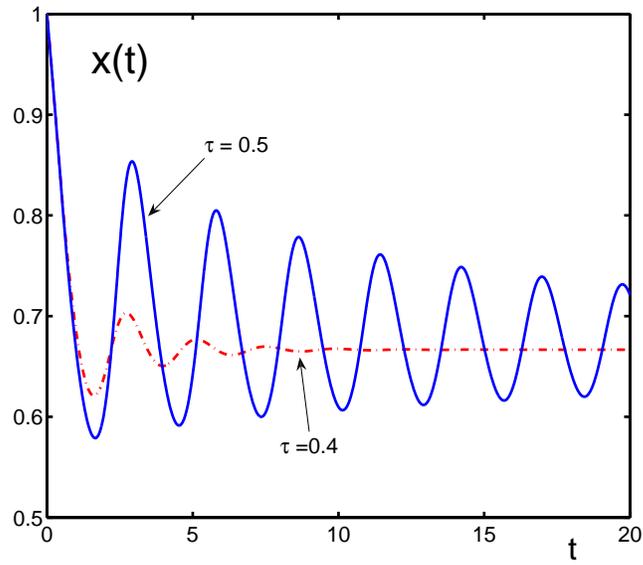}}
\caption{Solutions $x(t)$ to Eq. (\ref{60}) as functions of time
for the parameters $\tau=0.4$ (solid line) and $\tau=0.5$
(dashed-dotted line), where $\tau<\tau_0=0.505951$, with $\tau_0$
defined by (\ref{65}). Other parameters are: $x_0=1$, and $b=-2.5$.
The solutions $x(t)$ converge by oscillating towards their stationary point
$x_2^*=-1/(1+b)=2/3$ as $t\ra\infty$.}
\label{fig:Fig.14}
\end{figure}

\newpage
\begin{figure}[ht]
\centerline{\includegraphics[width=10cm]{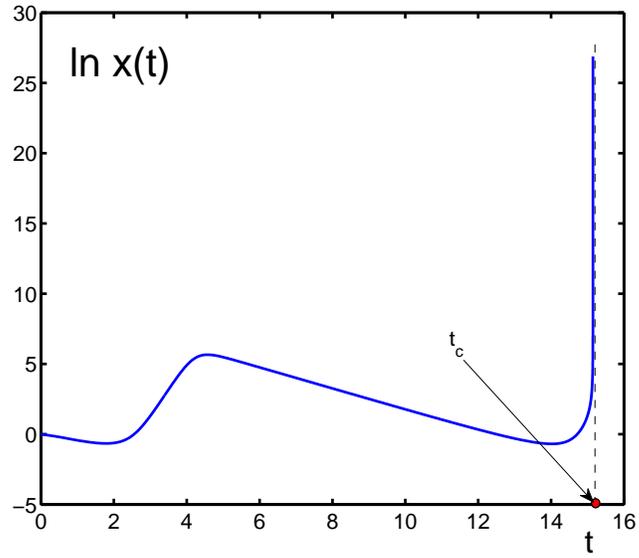}}
\caption{Behavior of the solution $x(t)$  in logarithmic scale to Eq. (\ref{60})
as a function of time for the parameters $b=-2.5$, $\tau=0.6272$, and
the initial history $x_0=1$. The solution $x(t)\ra\infty$ as
$t\ra t_c=15.1498$, where $t_c$ is defined by (\ref{70}).}
\label{fig:Fig.15}
\end{figure}

\newpage

\begin{figure}[ht]
\centerline{\includegraphics[width=10cm]{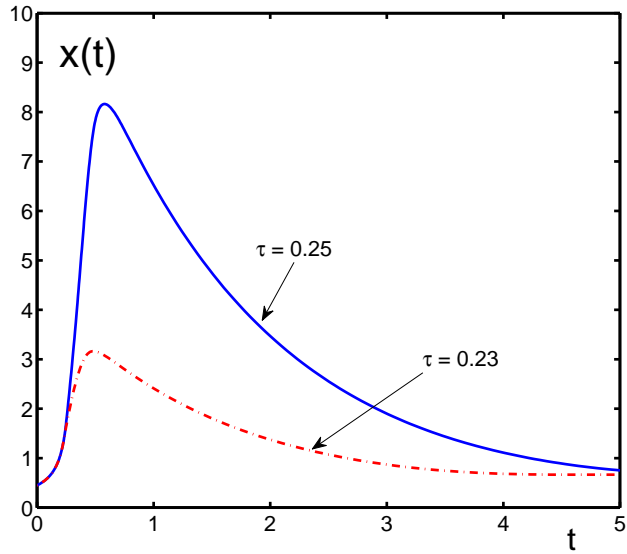}}
\caption{Temporal behavior of solutions $x(t)$ to Eq. (\ref{60})
as functions of time for the parameters $b=-2.5$ and $\tau=0.25$ (solid
line); $\tau=0.23$ (dashed-dotted line); and the initial history $x_0=1$.
The solution $x(t)\ra x^*=2/3$, when $t\ra\infty$.}
\label{fig:Fig.16}
\end{figure}

\newpage

\begin{figure}[ht]
\centerline{\includegraphics[width=10cm]{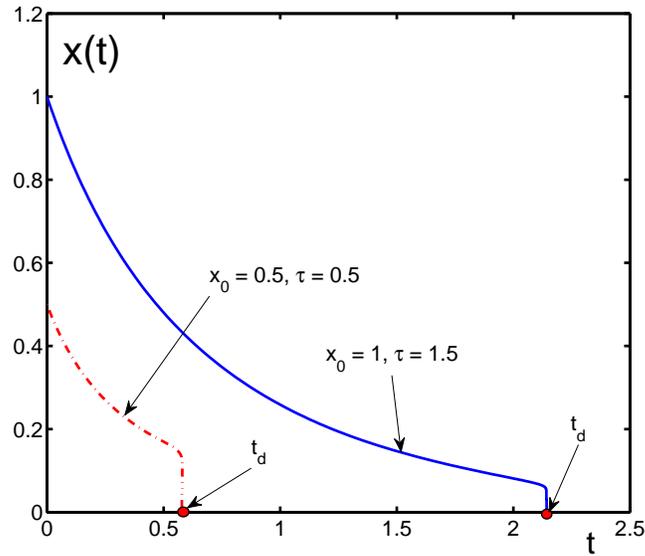}}
\caption{Behavior of solutions $x(t)$ to Eq. (\ref{75})
as functions of time for $b=-2.5$, lag $\tau=1.5$, history
$x_0=1$  (solid line), and lag $\tau=0.5$, history $x_0=0.5$
(dashed-dotted line). At the moment $t=t_d$, defined as
$1+bx(t_d-\tau)=0$, solutions monotonically decay to zero,
$x(t_d)=0$, $\dot{x}(t)|_{t=t_d}=-\infty$. The death time for
the population, represented by the solid line, is $t_d=2.14158$, 
and the death time for the population shown by the dashed-dotted
line, is $t_d=0.58004$.}
\label{fig:Fig.17}
\end{figure}

\newpage

\begin{figure}[ht]
\centerline{\includegraphics[width=10cm]{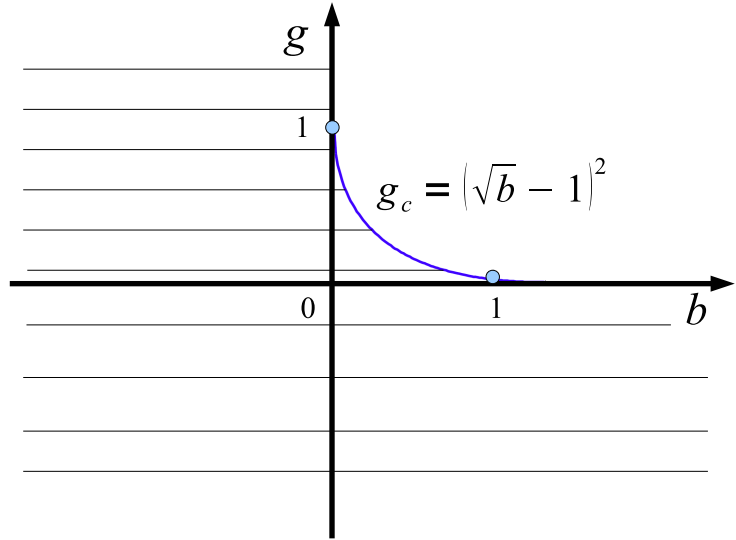}}
\caption{Region of stability (shaded) in the parameter plane
$b-g$ for the stationary solutions in the case of symbiosis with 
mutual interactions.}
\label{fig:Fig.18}
\end{figure}

\newpage

\begin{figure}[ht]
\vspace{9pt}
\centerline{
\hbox{ \includegraphics[width=7cm]{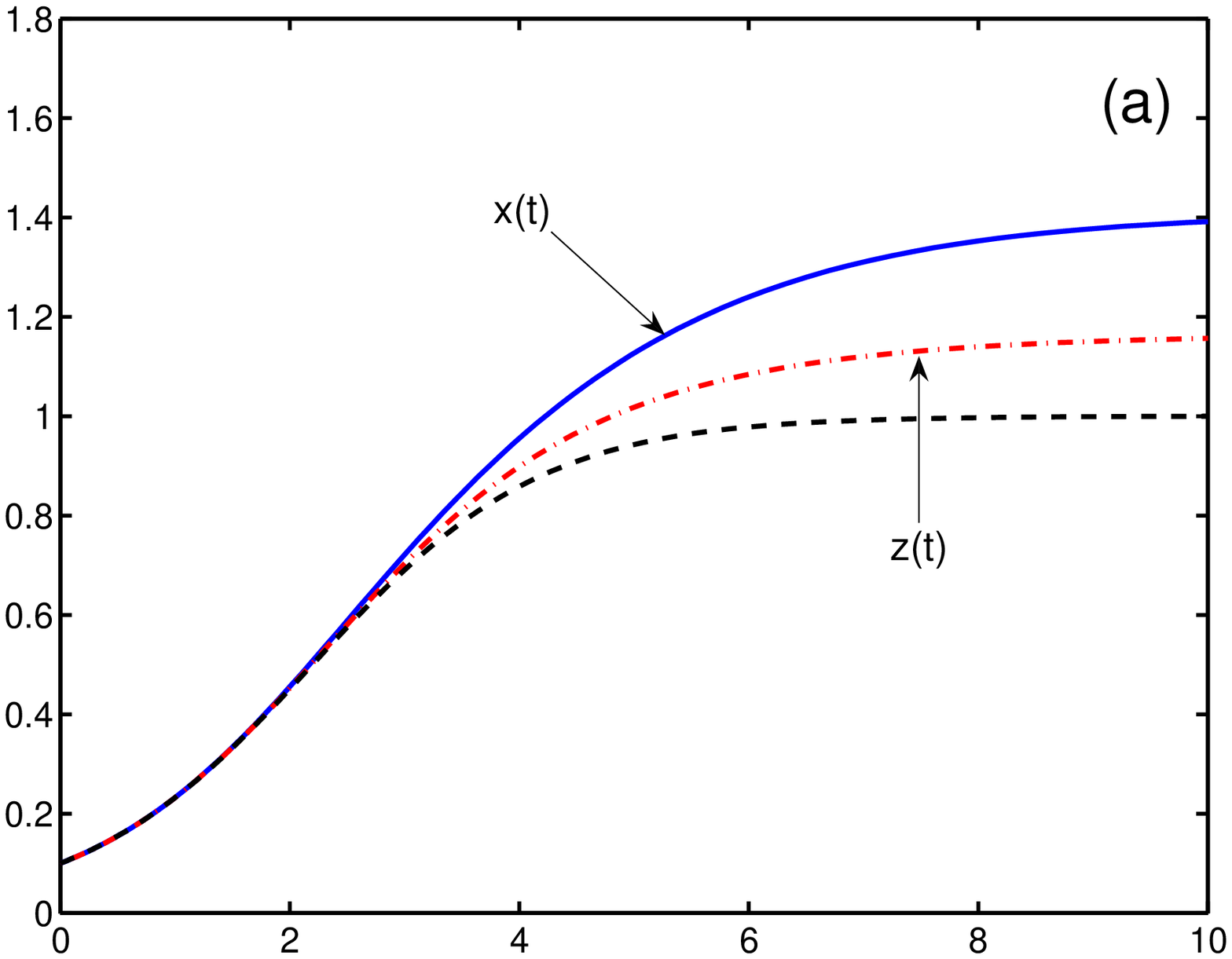} \hspace{2cm}
\includegraphics[width=7cm]{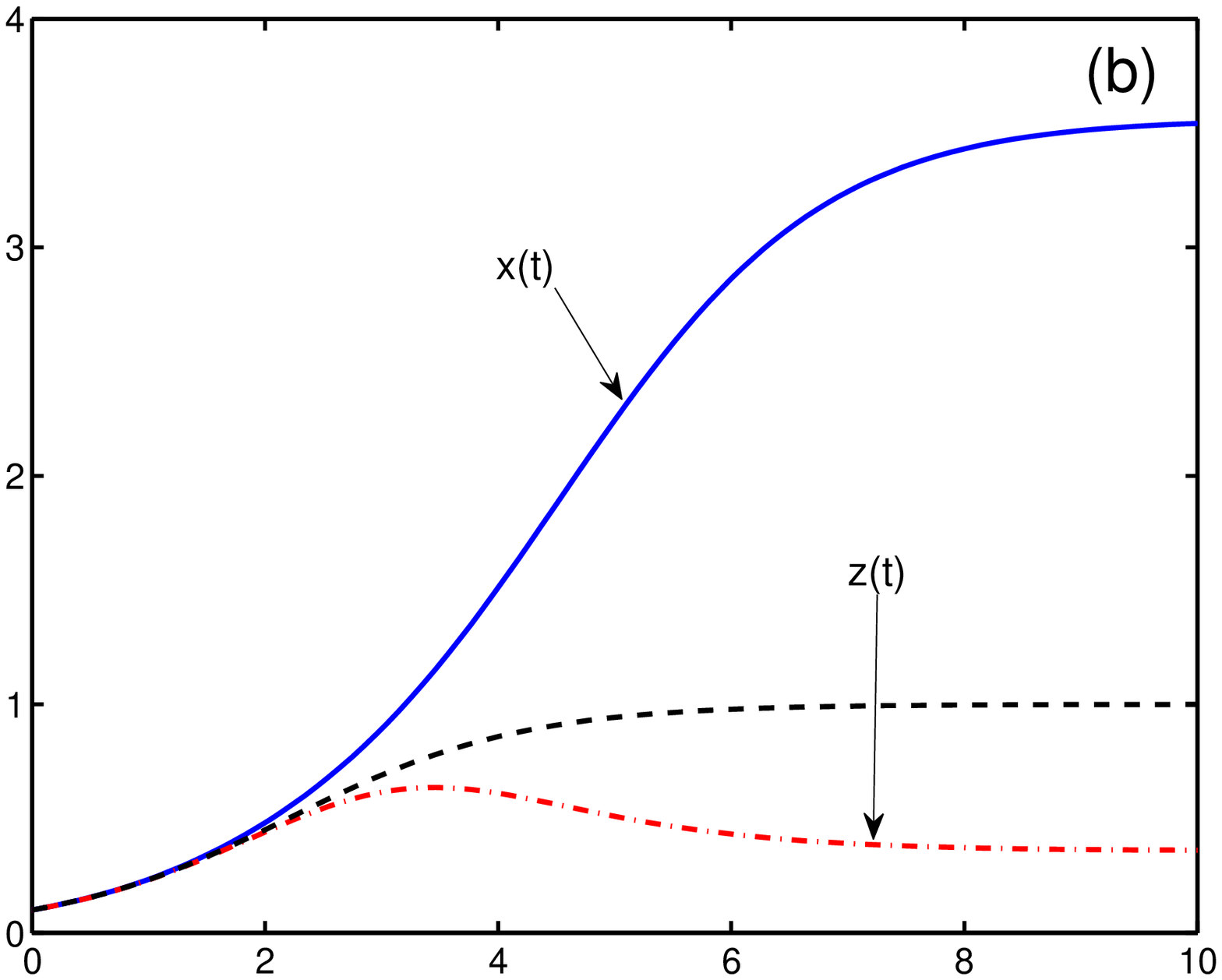} } }
\vspace{9pt}
\centerline{
\hbox{ \includegraphics[width=7cm]{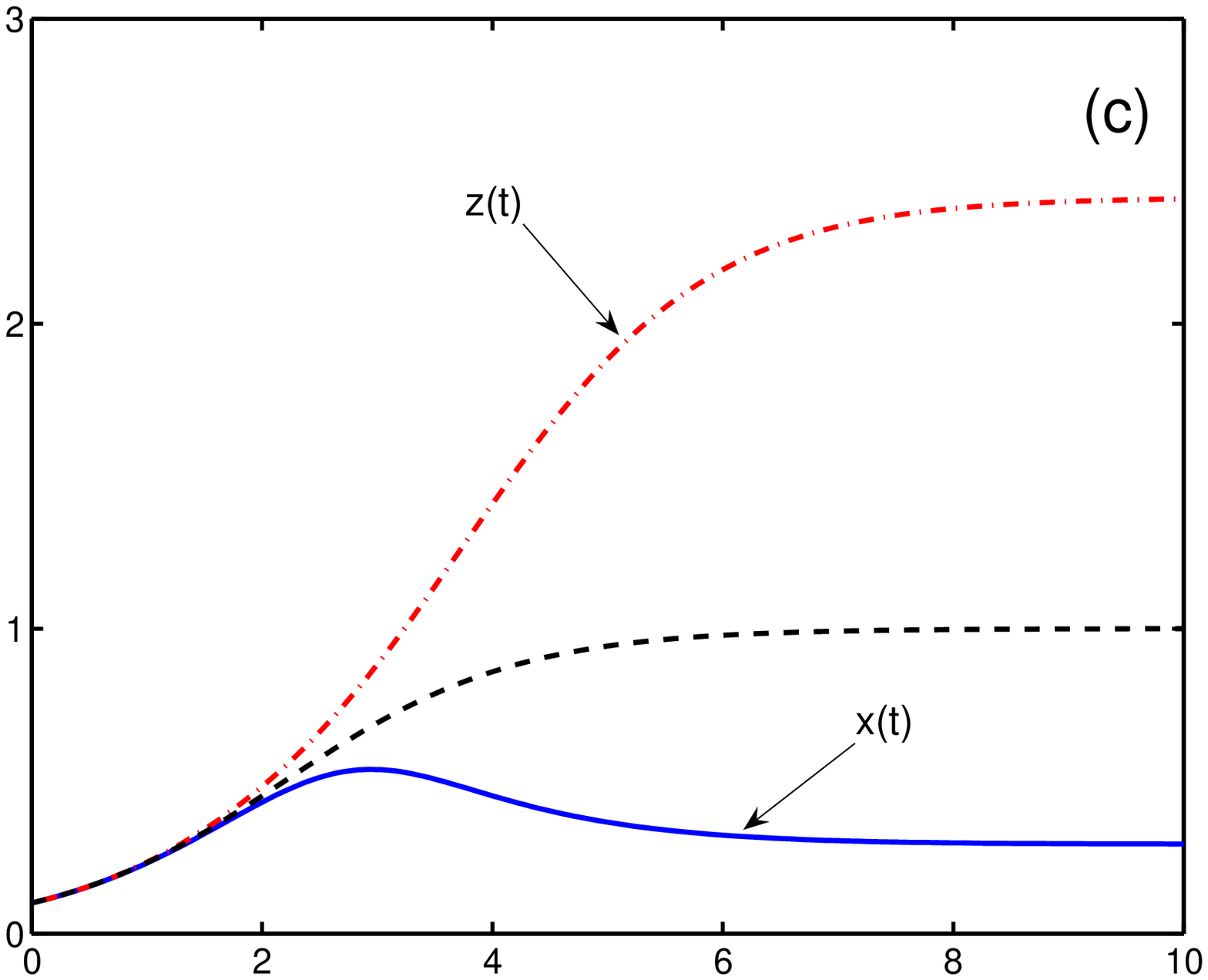} \hspace{2cm}
\includegraphics[width=7cm]{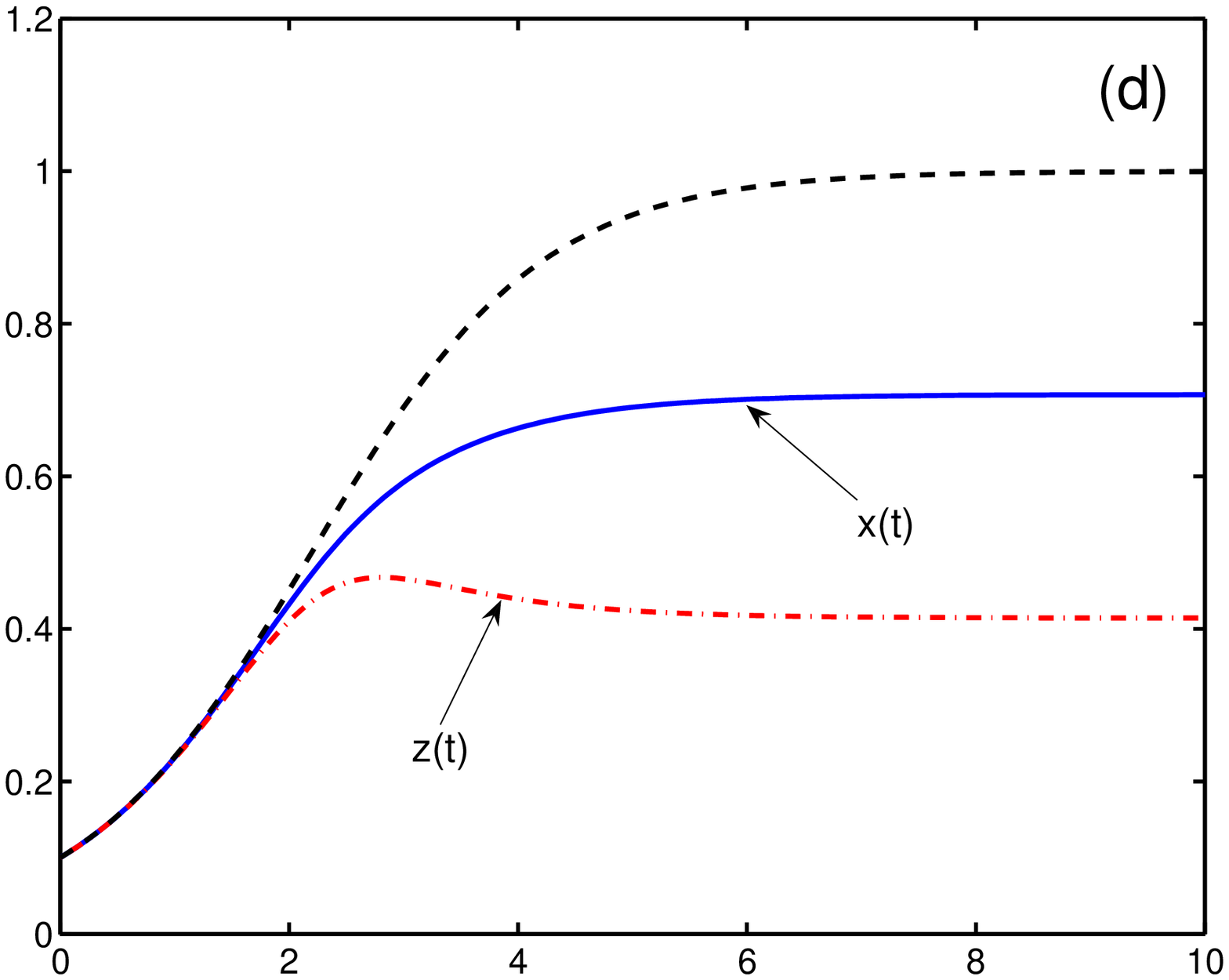} } }
\caption{Comparison of the symbiotic solutions $x(t)$ (solid
line) and $z(t)$ (dashed-dotted line), for the symbiosis
with mutual interactions (\ref{100}), with the solutions
$x(t) = z(t)$ (dashed line) of the decoupled equations (\ref{107})
for the same initial conditions $x_0=z_0=0.1 < 1$, with different 
symbiotic parameters $b$ and $g$: (a) $b = 0.25$, $g = 0.1 < g_c = 0.25$, 
the stationary points of the symbiotic equations being $x^* = 1.411$,
$z^* = 1.164$; (b) $b = 2$, $g = -0.5$, the fixed points of the 
symbiotic equations being $x^*=3.562$, $z^*=0.360$; (c) $b = -1$, $g = 2$, 
the symbiotic fixed points being $x^* = 0.293$, $z^* = 2.414$; 
(d) $b = -1$, $g = -2$, the symbiotic fixed points being $x^* = 0.707$,
$z^* = 0.414$.}
\label{fig:Fig.19}
\end{figure}

\newpage

\begin{figure}[ht]
\centerline{\includegraphics[width=10cm]{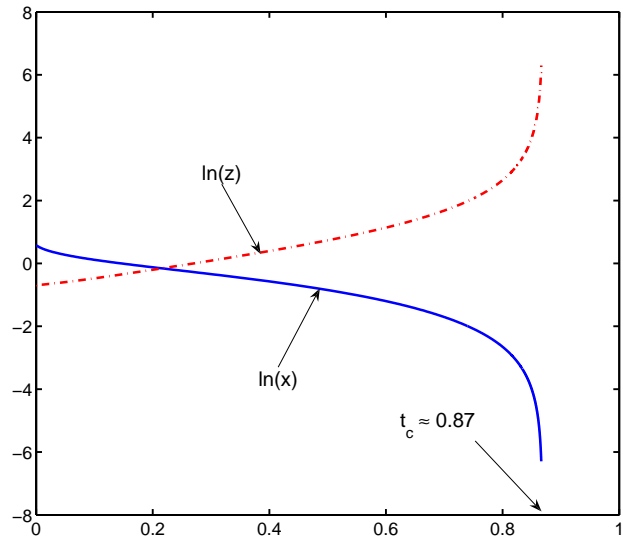}}
\caption{Behavior of the solutions  in logarithmic scale to equations (\ref{100})
in the case of finite-time death and singularity, $x(t)$ (solid line)
and $z(t)$ (dashed-dotted line), for the parasitic relations with the
symbiotic coefficients $b = -1$, $g = -2$, under the initial conditions
$x_0 = 1.8$, $z_0 = 0.5$. For these parameters, the critical time
is $t_c = 0.87$.}
\label{fig:Fig.20}
\end{figure}

\newpage

\begin{figure}[ht]
\centerline{\includegraphics[width=10cm]{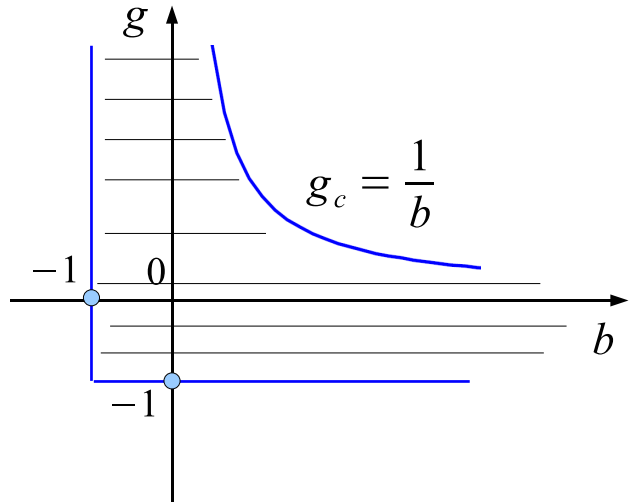}}
\caption{Region of stability (shaded) in the parameter plane
$b-g$ for the fixed points in the case of symbiosis without direct
interactions.}
\label{fig:Fig.21}
\end{figure}

\newpage

\begin{figure}[ht]
\vspace{9pt}
\centerline{
\hbox{\includegraphics[width=7cm]{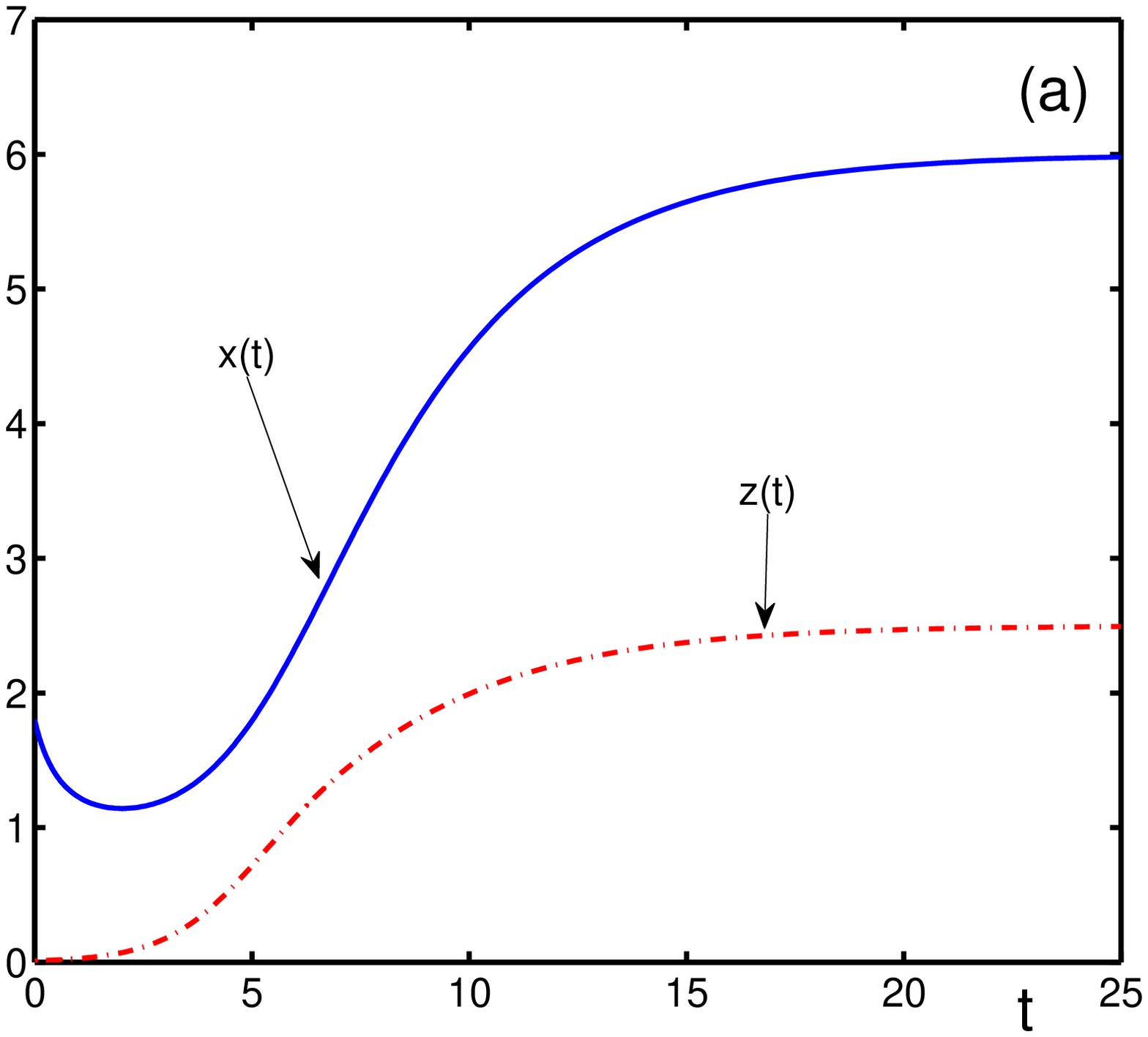} \hspace{2cm}
\includegraphics[width=7cm]{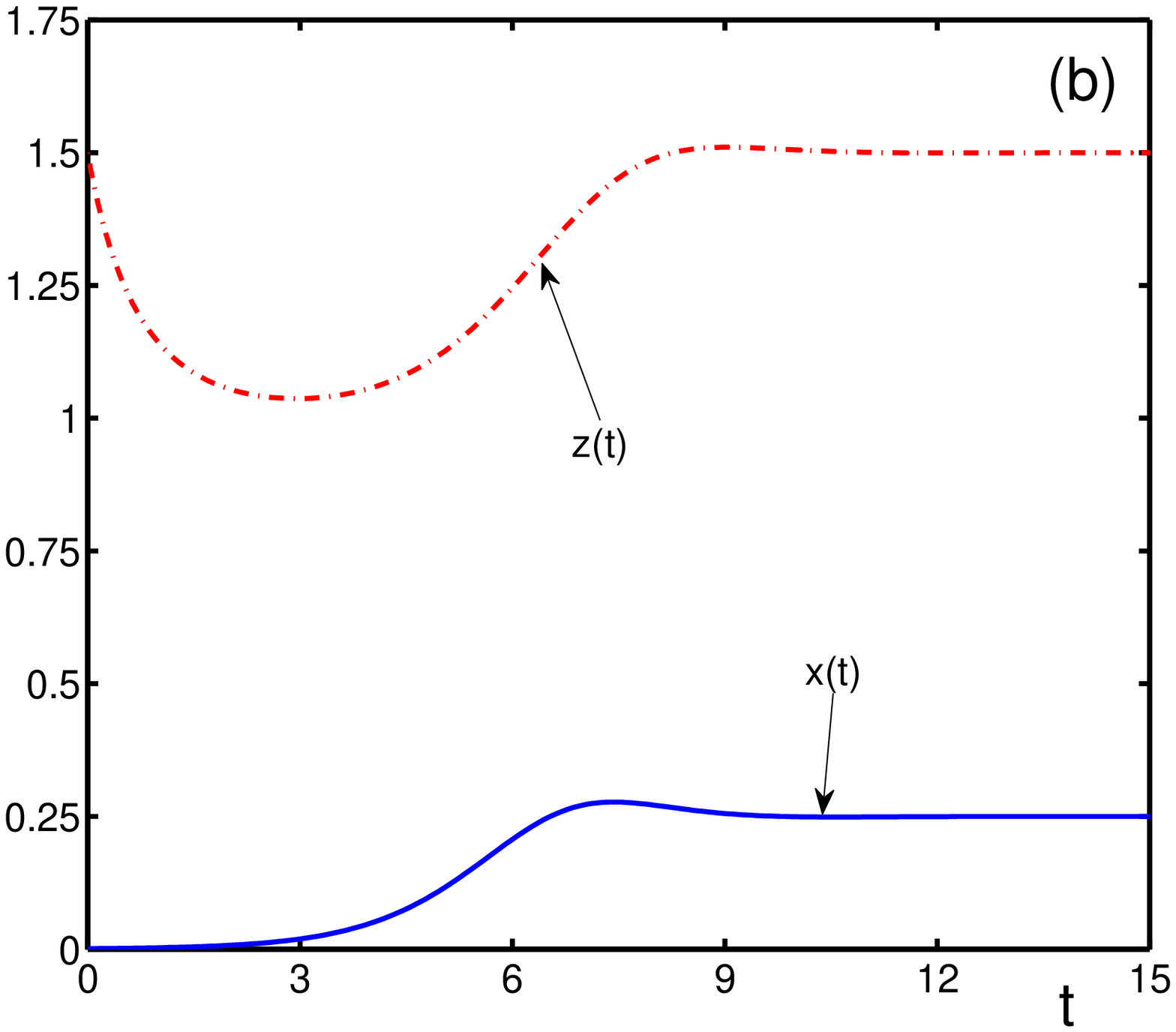} } }
\caption{Nonmonotonic convergence to stationary states of $x(t)$ 
(solid line) and $z(t)$ (dashed-dotted line)  as functions
of time, in the case of symbiosis without direct interactions 
(\ref{116}): (a) for the initial conditions $x_0=1.8$, $z_0=0.01$ 
and the parameters $b=2$, $g=0.25$. The functions $x(t)\ra x^*$ and 
$z(t)\ra z^*$, when $t\ra\infty$, the fixed points being $x^* =6$, 
$z^* =2.5$; (b) for the initial conditions $x_0=0.001$, $z_0=1.5$,
and the parameters $b=-0.5$, $g=2$. Functions $x(t)\ra x^*=0.25$ and 
$z(t)\ra z^*=1.5$, when $t\ra\infty$.}
\label{fig:Fig.22}
\end{figure}

\newpage

\begin{figure}[ht]
\centerline{\includegraphics[width=10cm]{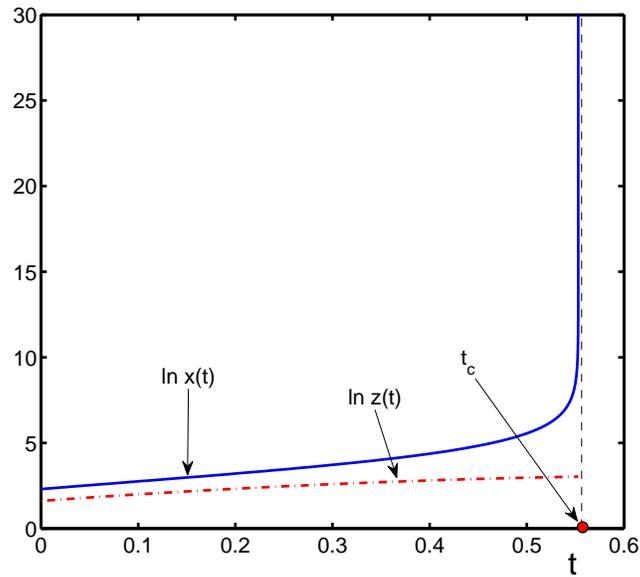}}
\caption{Behavior of the solutions  $x(t)$ (solid line) 
and $z(t)$ (dashed-dotted line)  in logarithmic scales in the case of 
symbiosis without direct interactions (\ref{116}) in the presence of 
the finite-time singularity for the symbiotic coefficients 
$b =-0.75$, $g =-0.25$, and the initial conditions $x_0 =10$, $z_0 =5$. 
The critical time is $t_c = 0.55303$.}
\label{fig:Fig.23}
\end{figure}

\newpage

\begin{figure}[ht]
\vspace{9pt}
\centerline{
\hbox{ \includegraphics[width=7cm]{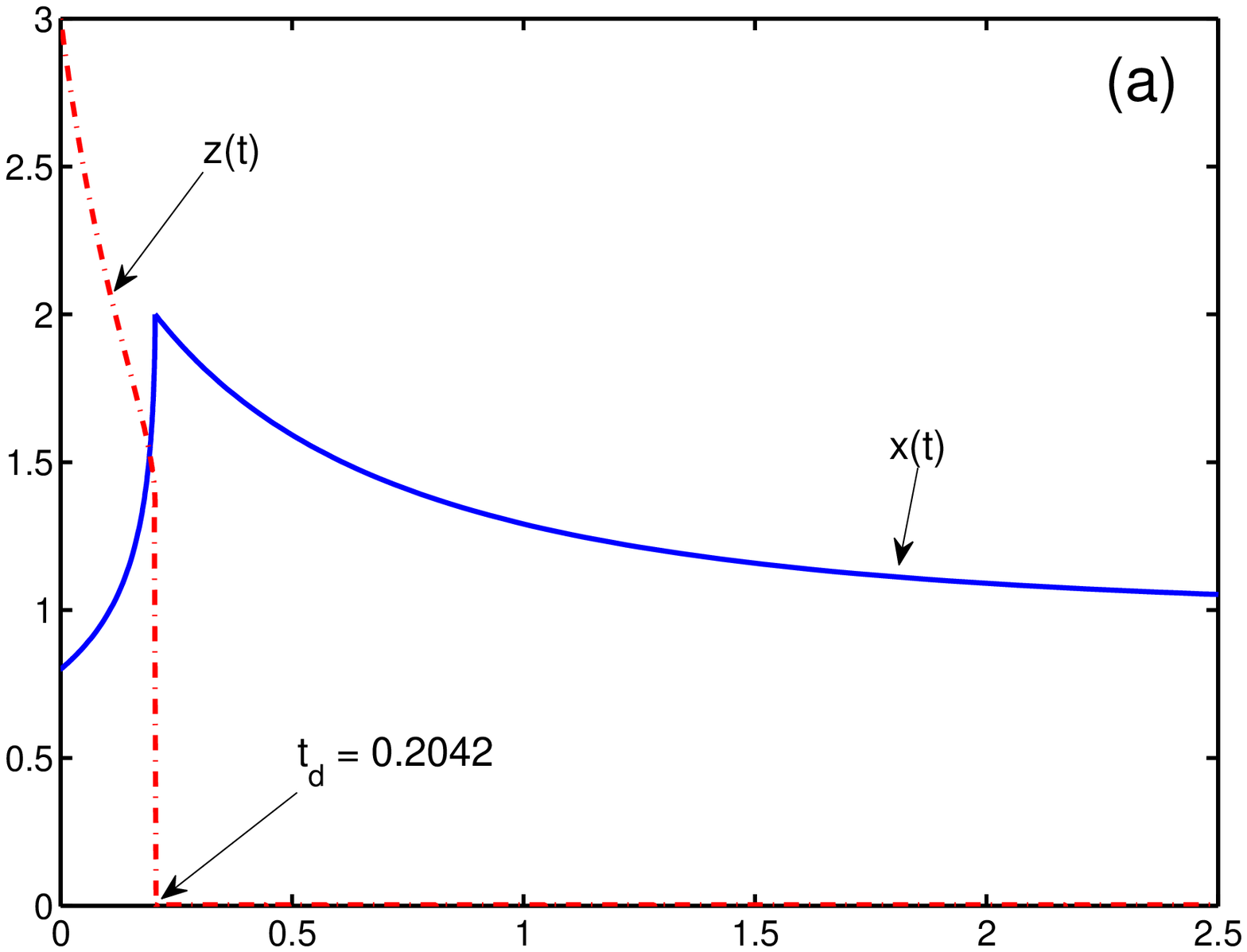} \hspace{2cm}
\includegraphics[width=7cm]{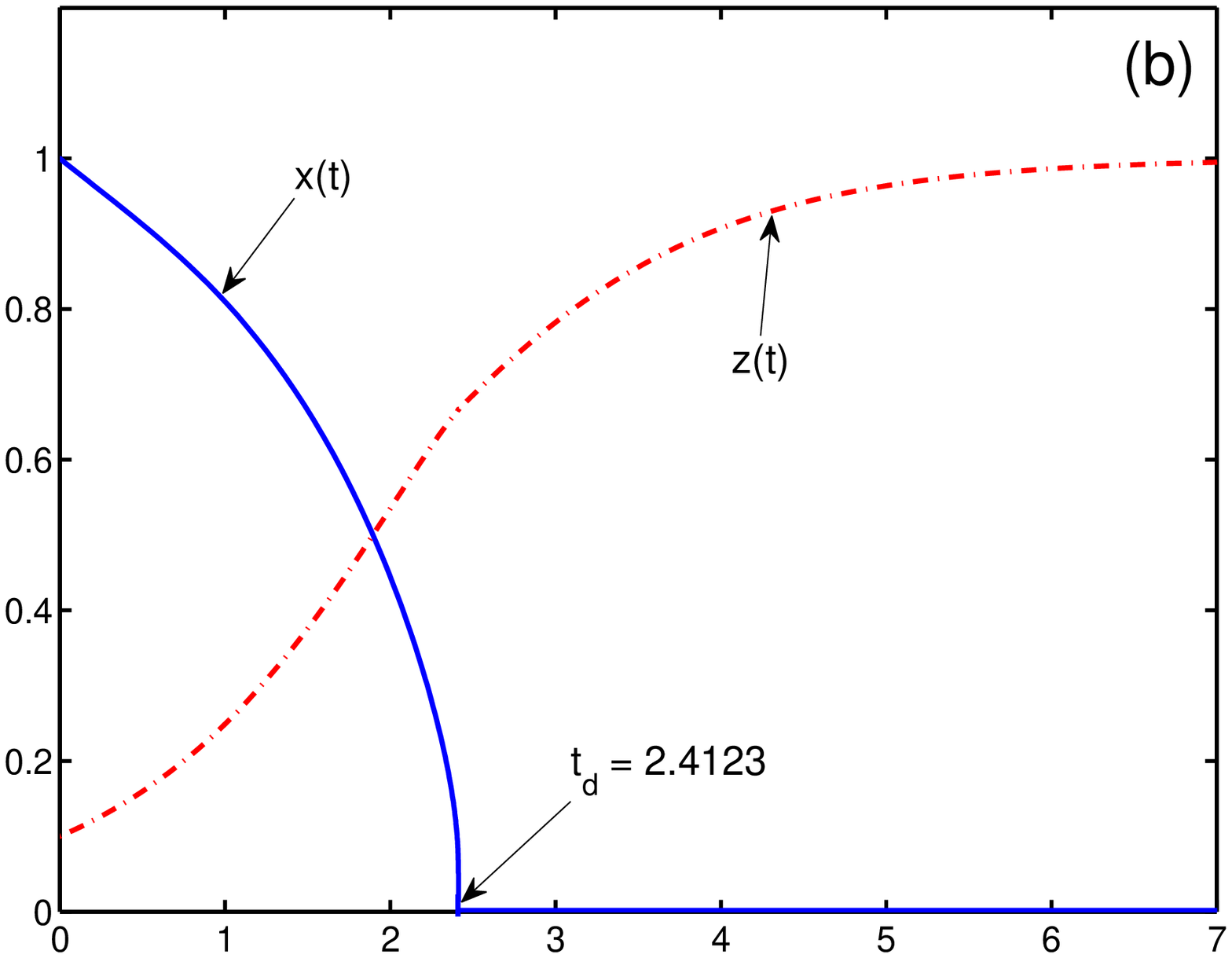} } }
\caption{Finite-time death in the case of symbiosis without
direct interactions. Temporal behavior of solutions $x(t)$ (solid
line) and $z(t)$ (dashed-dotted line) for different symbiotic
parameters and initial conditions: (a) $b = -0.75$, $g = -0.5$,
$x_0 = 0.8$, $z_0 = 3$, the death time being $t_d = 0.204$;
(b) $b = -1.5$, $g = 1$, $x_0 = 1$, $z_0 = 0.1$, with the death
time $t_d = 2.412$.}
\label{fig:Fig.24}
\end{figure}

\end{document}